\documentclass[twocolumn, usenatbib]{mnras}
\usepackage{amsthm}
\usepackage[utf8]{inputenc}
\usepackage{amssymb}
\usepackage{graphicx}
\usepackage{amsmath}
\usepackage{float}
\newtheorem*{theorem}{Theorem}

\theoremstyle{definition}

\usepackage{verbatim}
\usepackage{hyperref}

\newcommand{\eps}{$\varepsilon$\textsc{ppsilon}}

\title[Gaussian Error Distributions in 21-cm Power Spectra]{Why and When to Expect Gaussian Error Distributions in Epoch of Reionization 21-cm Power Spectrum Measurements}
\author[M. J. Wilensky, J. Brown, and B. J. Hazelton]{Michael J. Wilensky,$^{1}$\thanks{michael.wilensky@manchester.ac.uk} Jordan Brown,$^{2}$ Bryna J. Hazelton$^{3, 4}$
\\
$^{1}$Jodrell Bank Centre for Astrophysics, University of Manchester, Manchester M13 9PL, UK\\
$^{2}$Group in Logic and the Methodology of Science, University of California Berkeley, Berkeley, CA 94720, USA\\
$^{3}$Department of Physics, University of Washington, Seattle, WA 98195, USA\\
$^{4}$eScience Institute, University of Washington, Seattle, WA 98195, USA
}

\begin{document}

\maketitle

\begin{abstract}
    We explore error distributions in Epoch of Reionization 21-cm power spectrum estimators using a combination of mathematical analysis and numerical simulations. We provide closed form solutions for the error distributions of individual bins in 3d-power spectra for two estimators currently in use in the field, which we designate as ``straight-square" and ``cross-multiply" estimators. We then demonstrate when the corresponding spherically binned power spectra should (and should not) have Gaussian error distributions, which requires appealing to nonstandard statements of the central limit theorem. This has important implications for how upper limits are reported, as well as how cosmological inferences are performed based on power spectrum measurements. Specifically, assuming a Gaussian error distribution can over or underestimate the upper limit depending on the type of estimator, and produces overly compact likelihood functions for the power spectrum.
\end{abstract}


\section{Introduction}

The Epoch of Reionization (EoR), a period when the Hydrogen content of the universe went from being predominantly neutral to being predominantly ionized, can in principle be mapped directly using the 21-cm line from neutral Hydrogen \citep{Furl2006, Morales2010}. Since the 21-cm line arises from a forbidden transition in neutral Hydrogen, it acts as a direct tracer of the progress of cosmic reionization. Until imaging sensitivity can be achieved, observational experiments focus on constraining the power spectrum of this cosmic reionization signal i.e. the Fourier dual of the two-point correlation function of the map. This is done using radio interferometers, a few of which have produced steadily deepening upper limits, such as the Precision Array for Probing the Epoch of Reionization \citep[PAPER;][]{Parsons2010}, the Giant Metrewave Radio Telescope \citep[GMRT;][]{Paciga2013}, the Murchison Widefield Array \citep[MWA;][]{Tingay2013, Wayth2018}, the LOw Frequency ARray \citep[LOFAR][]{vanHaarlem2013}, and the Hydrogen Epoch of Reionization Array \citep[HERA;][]{DeBoer2017}. The layout of the interferometric baselines determine the sensitivity of the array to transverse spatial (angular) modes, while its frequency structure determines line-of-sight spatial characteristics when searching for a redshifted line such as the cosmic reionization signal. Current power spectrum estimators generally fall into two camps: delay spectra, which directly Fourier transform interferometric visibilities for each baseline along the frequency axis, and imaging power spectra, which grid the visibilities before Fourier transforming \citep{Parsons2012, Liu2014a, Morales2019}. Either process generates a power spectrum in 3-dimensional $\bold{k}$-space. The quantity most readily compared to theoretical predictions is the spherically averaged power spectrum (i.e. averaging the power spectrum in spherical shells in the aforementioned $\bold{k}$-space). Since the target signal is extremely faint relative to other sources of radio emission, astrophysical or otherwise, current experiments generally focus on systematic mitigation and subsequent placement of an upper limit on the strength of the power spectrum signal on various spatial scales.

As upper limits on the cosmological 21-cm power spectrum signal from the Epoch of Reionization push closer to detection, it will be increasingly important to understand the exact statistical properties of the measurement. Though estimators vary, the power spectrum is inherently a quadratic statistic, and this can make tracking the exact sampling distributions of the estimators complicated. As such, there exist a number of noise characterization techniques meant to provide an estimate of the size of thermal fluctuations in the measurement, often without explicit reference to these sampling distributions. 

For a thorough exploration of noise estimation techniques in delay spectrum analysis, see \citet{Tan2021}. Such noise estimation techniques apply directly to power spectrum estimates formed by PAPER and HERA. These techniques include bootstrapping, visibility differencing, analytic noise power spectrum estimation \citep{Liu2020}, and covariance matrix propagation. Additionally, the LOFAR EoR power spectrum pipeline has deployed a combination of propagation of Stokes V information and Gaussian Process Regression \citep[GPR;][]{Mertens2018, Kern2021} to inform noise estimates \citep{Patil2017, Mertens2020}. Other end-to-end noise propagation techniques are deployed in MWA power spectrum estimates that have overlap with previously mentioned techniques here \citep{Trott2016b, Barry2019a}. 

While these noise estimates are all physically grounded and generally indicate the size of statistical uncertainties, a main point of \cite{Tan2021} is that the exact statistical question answered by each estimator varies between them. For instance, it has become a standard to reference the ``2$\sigma$" upper limit on the reionization signal, but if the sampling distribution is non-Gaussian, then a 2$\sigma$ upper limit made with reference to some estimated covariance will be different than one made by, say, a 97.7\% one-tailed confidence interval estimated by Monte Carlo simulation. An exact value of the upper limit calculated from the sampling distribution corresponding to the power spectrum estimator would be the clearest item to interpret. In some examples, it is explicitly assumed that the sampling distribution is Gaussian, e.g. \citet{Li2019}. In \citet{Tan2021}, the sampling distribution is checked somewhat exhaustively for approximate Gaussianity, and an exact sampling distribution is supplied for their estimator. This is explained by reference to the well-known central limit theorem (CLT). We remark that there are several variations of the CLT, not well-known to the community at large, and it is in fact the more nuanced variations of the CLT that are required to justify Gaussian error distributions in EoR 21-cm power spectrum measurements. The purpose of this work is to present and explore these more advanced variations in order to delineate when and why the assumption of Gaussianity is mathematically justified. 

The paper is laid out as follows. In \S\ref{sec:CLT} we formally present the variations of the CLT that are relevant to our formal arguments, as well as the relevant mathematical background to understand them. In \S\ref{sec:circ_gauss}, we refresh the reader on complex circular Gaussian random vectors. In \S\ref{sec:straight-square}-\ref{sec:cross-mult}, we discuss the sampling distributions of the thermal noise\footnote{Strictly speaking, the noise term in the interferometric visibilities receives contribution both from thermal fluctuations in the instrument, statistical fluctuations in the emission of non-thermal sources, and other effects. For high-redshift 21-cm observations, the noise fluctuations are dominated by non-thermal galactic synchrotron radiation. However, it is a historical convention to refer to the collective noise fluctuations as ``thermal noise" despite non-thermal contributions. The connection is that, at each frequency, the brightness of a non-thermal source is identified with a blackbody of corresponding temperature so that its contribution to the system temperature (called its brightness temperature) predicts the appropriate noise fluctuations per the radiometer equation. See e.g. \citet{bennett2003} and \citet{deOliveira-Costa2008}.} in 3-d power spectra, and explore the effects of incoherent averaging (i.e. averaging of power spectra, rather than visibilities) assuming all samples are independent. We broadly classify these into ``straight-square" estimators, e.g. \citet{Patil2017} and \citet{Mertens2020}, and ``cross-multiply" estimators, such as in \citet{HERA2022, HERA2022C} and \citet{Barry2019b}. We analytically calculate sampling distributions for both, and then explore how each type of distribution can (but might not) converge to a Gaussian via different averaging schemes. In the latter case, we reproduce the averaging schemes used on actual data. In \S\ref{sec:dep_var}, we explore the effects of combining statistically dependent samples. We summarize our results and draw our conclusions in \S\ref{sec:conc}.

\section{Central Limit Theorem}
\label{sec:CLT}
In this section, we review the classical version of the central limit theorem and give generalizations applicable to the error distributions in 21cm measurement, as the hypotheses of the classical theorem are not satisfied by the sums of error terms arising in current 21cm measurement processing pipelines. Central limit theorems are about convergence of an infinite sequence and therefore do not apply directly to any finite sum of random variables. Nonetheless, they provide a theoretical justification for modeling such sums as having a Gaussian distribution in many circumstances. We give a precise statement about finite sums from \citet{Shevtsova2021} after a short exposition regarding the infinite limit.
\\ \\
The classical central limit theorem states that the finite sums of an infinite sequence of independent and identically distributed random variables converges in distribution to a Gaussian. The conclusion is not necessarily true if either the assumption of independence or the assumption of identical distribution is dropped. A sequence of random variables, $(X_1, X_2, X_3, ...)$, is identically distributed if, for any $X_i$ and $X_j$ in the sequence, the cumulative distribution function of $X_i$ is equal to the cumulative distribution function of $X_j,$ i.e. $F_{X_i} = F_{X_j}.$ A sequence is independent if, for any finite set $I$ of natural numbers and finite sequence $(a_i)_{i\in I},$ we have $\mathbb{P}(i\in I\rightarrow X_i>a_i)=\prod_{i\in I} \mathbb{P}(X_i>a_i)$, where $\mathbb{P}$ denotes probability of an event. If the variables have probability density functions, this is equivalent to the pdf of the joint distribution being given pointwise by the product of the pdfs of the individual variables. Note that uncorrelated variables may still be dependent; this is the case for $X$ and $|X|$ whenever $X$ has a pdf that is an even function. Additionally, sequences of pairwise independent random variables may not be independent sequences. The hypotheses for the classical central limit theorem are, taken together, quite stringent. We give statements from the exposition in \citet{billingsley2012probability}.
\begin{theorem}[\citet{billingsley2012probability} Theorem 27.1]
Suppose that $\{X_n\}$ is an independent sequence of random variables having the same distribution with mean $c$ and finite positive variance $\sigma^2.$ If $S_n = X_1+...+X_n,$ then $\frac{S_n-nc}{\sigma\sqrt{n}}\rightarrow N(0,1)$.
\end{theorem}
Here, $N(0,1)$ indicates a standard normal random variable. In an extreme case of dependency where all variables $X_i$ in the sequence are equal, $\frac{S_n-nc}{\sigma\sqrt{n}}$ is a (zero-mean) translation of $X_i$ and does not converge to a Gaussian if $X_i$ is not itself Gaussian. Most pipelines combine non-independent samples in some form or another, e.g. when computing spherical power spectra from 3d-power spectra, or when averaging cross power spectra over baseline-pairs that share a baseline \citep{Tan2021}. This leads to dependency between noise measurements, so this hypothesis is not strictly satisfied.
\\ \\
In addition, the random noise variables are not identically distributed. We see in \S\S\ref{sec:straight-square}-\ref{sec:cross-mult} that the relevant variables are Exponential or Laplacian, as they arise from a squaring or product of circular complex Gaussians, and that their scale parameters vary. If successively smaller scale parameters appear in the sum, the sum may never look Gaussian, with approximately exponential tails. However, in many cases one does obtain Gaussian distributions \S\S\ref{sec:straight-square}-\ref{sec:dep_var}. The case of independent, but non-identically distributed random variables, is characterized by the Lyapunov CLT:
\begin{theorem}[\citet{billingsley2012probability} Theorem 27.3]
Suppose that $\{X_n\}$ is an independent sequence of random variables, each with mean zero and finite variance $\sigma_n^2,$ and there is some $\delta>0$ such that, with $s_n^2=\sum_{i=1}^n \sigma_i^2,$ $\lim_{n\rightarrow\infty}\sum_{i=1}^n s_n^{-(2+\delta)}\mathbb{E}[|X_i|^{2+\delta}]=0.$ Then $S_n/s_n\rightarrow N(0,1).$
\end{theorem}
There are also effective forms of the Lyapunov central limit theorem, i.e. forms of the theorem that bound the rate of convergence. We state the sharpest bound we know of on the difference between the cdfs and a Gaussian cdf.\footnote{This theorem makes use of the \textit{supremum}, which for a bounded set of real numbers is the lowest upper bound of the set. For example, the supremum of the real open interval $(-1, 1)$ is 1.}
\begin{theorem}[\citet{Shevtsova2021} Theorem 1]\label{Berry-Essen}
Suppose that $\{X_n\}$ is an independent sequence of random variables, each with mean zero, $E[X_n^2]=\sigma_n^2,$ and $E[|X_n|^3]=\beta_n$, with $\sigma_n,\beta_n$ finite and $\sum_{i=1}^n\sigma_i^2=1$. Writing $F_n$ for the cdf of $\sum_{i=1}^n X_i$ and $\Phi$ for the cdf of $N(0,1)$, we have
$$\sup_{x\in\mathbf{R}}|F_n(x)-\Phi(x)|\leq 0.56\sum_{i=0}^n \beta_i$$
\end{theorem} 
The quantity on the left-hand-side measures the distance of the cdf of the $n$th partial sum from a Gaussian cdf. The right-hand-side is typically easy to compute. If the left-hand-side goes to 0 in the infinite limit, then the sequence converges in distribution to a standard normal random variable. Therefore, the right-hand-side tells us how close a partial sum is to convergence at any given $n$.

Unfortunately, the random sequences at issue for the 21cm error bars are not always independent and the Lyapunov theorem does not strictly apply. However, some estimators may be constructed to satisfy a weak form of independence, introduced in \citep{HoeffRobbins}, known as $m-$dependence for a fixed natural number $m$. Essentially, if the statistical dependence only has a fixed finite range among the samples and the Lyapunov condition is satisfied, then there is convergence to Gaussianity. A sequence $(X_i)_{i=0}^\infty$ of random variables is said to be \textit{$m$-dependent} if, whenever $s-r>m,$ $(X_0,X_1,...,X_r)$ and $(X_s,X_{s+1},...,X_n)$ are independent vector-valued random variables. That is, if we write $\mathbf{Y}_0=(X_0,X_1,...,X_r)$ and $\mathbf{Y}_1=(X_s,X_{s+1},...,X_n),$ then, for any rectangles $R_0=(a_0,b_0)\times...\times (a_r,b_r)$ and $R_1=(a_s,b_s)\times...\times(a_n,b_n),$ we have $\mathbb{P}(\mathbf{Y}_0\in R_0\textrm{ \& }\mathbf{Y}_1\in R_1)=\mathbb{P}(\mathbf{Y}_0\in R_0)\mathbb{P}(\mathbf{Y}_1\in R_1).$
\\ \\
If $(X_i)$ is an $m-$dependent sequence of random variables, we define, for each natural number $i,$ $A_i=\mathbb{E}[X_{i+m}^2]+2\sum_{j=1}^m\mathbb{E}[X_{i+m-j}X_{i+m}].$\footnote{Note that, if the sequence is $m-$dependent and $m\leq n,$ then the sequence is $n-$dependent, and we may obtain different $A_i$ depending on whether we consider the sequence as $m-$dependent or $n-$dependent; having fixed an $m$ for which we consider $(X_i)$ to be $m-$dependent at the beginning, the definition of the $A_i$'s is unambiguous.}
\begin{theorem}[\citet{HoeffRobbins} Theorem 1]
If $(X_i)$ is an $m-$dependent sequence of random variables, each with zero mean, and there is a real number $M$ such that, for all $i,$ $\mathbb{E}[|X_i|^3]\leq M$, and $\lim_{p\rightarrow\infty}p^{-1}\sum_{h=1}^p A_{i+h}$ exists uniformly for all $i,$ with limit $A,$ then the distribution function of $n^{-1/2}A^{-1}\sum_1^n X_i$ converges pointwise to the distribution function of $N(0,1).$

\end{theorem}
Hoeffding and Robbins's theorem shows that local dependency does not interfere with the central limit theorem, regardless of whether the random variables are identically distributed. It therefore theoretically justifies the assumption of approximate Gaussianity for the finite sums of random variables that arise in error estimates.

\section{Complex Circular Gaussian Random Vectors}
\label{sec:circ_gauss}

In this section, we quickly review some concepts of circular Gaussian random vectors and explain their relevance to the problem. We draw mainly from \cite{Gallager2013}, where detailed proofs of the following claims are held.

The visibility noise of a given baseline at each frequency, time is complex circular Gaussian \citep[i.e. the real and imaginary components are independent and identically distributed Gaussian random variables;][]{Thompson2017}, and is independent of the noise of every other baseline, frequency, time. This means the noise samples of all the visibilities may be considered together as jointly circular Gaussian. We explore particular estimators in more detail in the following sections, but all of them involve applying linear transformations to the visibilities to produce processed data whose noise-like component is a circular complex Gaussian random vector. These are then squared or cross-multiplied in a particular manner to produce the power spectrum estimator in question. Thus, if we can produce the probability density function of the square or cross multiplication of two independent circular complex Gaussian random variables and determine how such product distributions that are generally non-Gaussian behave under averaging, then we can see under what conditions a Gaussian error distribution is valid to assume.

Suppose the complex variables $(X_1, X_2, X_3 ..., X_N)$ are jointly Gaussian. Then we refer to the vector $\bold{X}$ whose $j$th component is $X_j$ as a Gaussian random vector. If 
\begin{equation}
    \begin{aligned}
    E[\bold{X}]=\boldsymbol{\mu} \\
    E[(\bold{X} - \boldsymbol{\mu})(\bold{X} - \boldsymbol{\mu})^\dag]= \bold{C} \\
    E[(\bold{X} - \boldsymbol{\mu})(\bold{X} - \boldsymbol{\mu})^T]= \boldsymbol{\Gamma} 
    \end{aligned}
    \label{eq:expec}
\end{equation}
where $\bold{X^T}$ and $\bold{X}^\dag$ are respectively the transpose and conjugate transpose of $\bold{X}$, then $\bold{X}$ is a complex Gaussian random vector variable with mean, $\boldsymbol{\mu}$, covariance matrix, $\bold{C}$, and pseudo-covariance matrix, $\boldsymbol{\Gamma}$. We write
%
\begin{equation}
    \bold{X} \sim \mathcal{CN}(\boldsymbol{\mu}, \bold{C}, \boldsymbol{\Gamma}).
\end{equation}
We say that $\bold{X}$ has circular symmetry if $\bold{X}e^{i\phi}$ is distributed identically to $\bold{X}$. That is, we can multiply the vector by an overall complex phase and the distribution is unaffected. Inspecting Equation \ref{eq:expec}, we see that it is necessary that
\begin{equation}
    \begin{aligned}
        \boldsymbol{\mu} = 0 \\
        \boldsymbol{\Gamma} = 0.
    \end{aligned}
\end{equation}
For a Gaussian random vector, this is also sufficient for circular symmetry, since the three expectations in Equation \ref{eq:expec} define all the statistical properties of the distribution.
From now on, we say that if $\bold{X}\sim\mathcal{CN}(0, \bold{C}, 0) \equiv \mathcal{CN}(0, \bold{C})$, then $\bold{X}$ is circular complex Gaussian. In this case, its probability density function, $f_\bold{X}(\bold{x} | \bold{C})$, is written\footnote{This form of the pdf may look unsightly to those who are only familiar with real-valued normal vectors, for which there is a factor of 2 in the exponent (as well as one associated with each factor of $\pi$), and a square root of the determinant. If one writes this circular Gaussian pdf in terms of the real and imaginary components of $\bold{X}$, so that it is instead describing a real-valued normal random vector of twice the length, then one finds these factors and operations in the appropriate locations (after an appropriate renaming of the covariance matrix as well). See \citet{Gallager2013}.}
\begin{equation}
    f_\bold{X}(\bold{x}) = \frac{\exp\big[-\bold{x}^\dag\bold{C}^{-1}\bold{x}\big]}{\pi^n\det\bold{C}}
\end{equation}
where $n$ is the number of components of $\bold{X}$. 

The most important property for circular complex Gaussian vectors for our purposes is that circular symmetry is maintained under linear transformations. That is, if
\begin{equation}
    \bold{Y} = \bold{A}\bold{X}
\end{equation}
where $\bold{A}$ is some constant matrix such that $\bold{A}\bold{C}\bold{A}^\dag$ is invertible, then
\begin{equation}
    \bold{Y} \sim \mathcal{CN}(0, \bold{A}\bold{C}\bold{A}^\dag).
\end{equation}
%
Note that if $\bold{A}\bold{C}\bold{A}^\dag$ is singular, then a dimensional reduction procedure is required to write the density in the form above, cf. Appendix A in \cite{Byrne2021}. Additionally, all linear functionals of circular complex Gaussian vectors are circular complex Gaussian variables, which is to say that for any (nonzero) vector $\bold{b}$, the scalar $\bold{b}^T\bold{X}$ is a circular complex random variable.

\section{``Straight-square" Power Spectrum Estimators}
\label{sec:straight-square}

We begin by describing power spectrum estimators that involve squaring the linearly transformed visiblities, such as that used in LOFAR upper limits. Denoting the linearly transformed visibilities as $\tilde{v}(\bold{k}, t)$, we may suppose that each voxel has a power of the form
\begin{multline}
    P(\bold{k}, t) = |\tilde{v}(\bold{k}, t)|^2 =  |s(\bold{k}, t) + n(\bold{k}, t)|^2 \\
    = |s(\bold{k}, t)|^2 + 2 \Re(s^*(\bold{k}, t)n(\bold{k}, t)) + |n(\bold{k}, t)|^2
    \label{eq:straight_square}
\end{multline}
where $s(\bold{k}, t)$ and $n(\bold{k}, t)$ are the signal and noise components of the linearly transformed visibilities, respectively, for the wave mode $\bold{k}$, at sidereal time $t$. Another data product for which this is relevant are the ``difference" cubes formed by Error Propagated Power Spectrum with Interleaved Observed Noise ($\varepsilon$\textsc{ppsilon}), which uses time-differenced data to form power-spectrum-esque objects \citep{Barry2019a}. These cubes are used both as a part of the $\varepsilon$\textsc{ppsilon} power spectrum estimator as well as by themselves as a noise metric, since they are sourced only by those entities that remain after time-differencing at the instrument cadence (typically thermal noise fluctuations and sometimes RFI sources, e.g. \citet{Wilensky2019}). In most cases it is fair to only consider the quadratic noise term in these difference cubes. $\varepsilon$\textsc{ppsilon} uses a generalization to the Lomb-Scargle periodogram \citep{VanderPlas2018} for power spectrum estimation, but its implementation can indeed be seen as first performing a linear transformation to the data, and then taking a sum of quadratic terms.

There are three terms in Equation \ref{eq:straight_square}, two of which contain contributions from the signal, one of which is a signal-noise cross term. It raises the question as to whether we should include this cross term into the description of the error statistics, since it clearly receives contribution from the noise. We note that this term is significant whenever the strength of the signal and noise terms are comparable, otherwise one of the two quadratic terms dominates. Without assuming some explicit model of the signal and the averaging scheme, it appears difficult to treat these terms to any satisfactory level of generality. 

All 21-cm power spectrum upper limits involve incoherent averaging (averaging after squaring or cross-multiplying, rather than \textit{coherent} averaging which takes place before), since at the very least samples at different $\bold{k}$-modes on the same spherical shell must be averaged together. Some power spectrum estimators also involve incoherently averaging different sidereal times. The incoherent averaging scheme is tied intimately to the treatment of the signal-noise cross term. For example, if one fixes a set of $\bold{k}$-bins, only ever observes one field on the sky (whose size is determined by the instrumental beam), the astrophysical foregrounds sources in that field are perfectly subtracted (or equivalently, no $\bold{k}$ bins affected by foregrounds are included), and no other systematics are present, then there is only one signal realization that ever enters the incoherent averaging scheme. Since we assume linearity in going from visibilities to $\tilde{v}(\bold{k}, t)$, we necessarily assume $n(\bold{k}, t)$ is a circular Gaussian random vector. An incoherent average (in this case the spherical average, since we have essentially assumed only one sidereal time) of the cross-terms then amounts to the real component of a weighted average of a subset of the complex components of this circular Gaussian random vector, and in this case the error distribution is potentially tractable since this term will have Gaussian distribution by the properties exposed in \S\ref{sec:circ_gauss}. However, one would still need to understand the signal variation between different $\bold{k}$-modes to obtain the appropriate sampling distribution. If more than one field is combined incoherently, then there are multiple realizations to contend with in the incoherent average, and then one must understand the nature of cosmic variance. Without a reliable model for variations in the signal from field to field or mode to mode, one could not propose a realistic accounting for the cross term. 

The presence of systematic effects tends to exacerbate the difficulty of the cross term. While astrophysical sources may behave similarly to the cosmological signal depending on the stationarity of the sources in the fields of interest, RFI is generally nonstationary, and has poorly understood statistics at faint intensities. Excess power from flagging \citep{Offringa2019a, Ewall-Wice2021, Wilensky2022} may also vary field-to-field and night-to-night, producing nontrivial statistics if not ameliorated, etc. Since a general treatment of systematics is beyond the scope of this work and we must assume something particular about the cosmological signal in order to characterize its effect in the cross term, we will ignore it in this analysis and focus ourselves on the noise-only term. This means this analysis is mainly applicable in the regime where the noise dominates the signal. Note that in the simplest case (only one field, no systematics other than potentially foregrounds), the Gaussianity of the distribution is unaffected by the addition of the cross term, though its mean and variance will be affected, which has important implications regarding reported confidence intervals, etc.

In general, there may be statistical dependency between different $\bold{k}$-modes or redundant measurements of the same $\bold{k}$-mode, and so the sampling distribution of the cylindrical or spherical power spectrum noise contamination is not necessarily straight-forward to compute. In general, if $\bold{X}$ is a circular Gaussian random vector (with arbitrary covariance), then the variable
\begin{equation}
    \bold{Y}' = \bold{X}^\dag\bold{B}\bold{X},
\end{equation}
for some complex matrix $\bold{B}$, has real and imaginary components that are a special case of ``Generalized $\chi^2$" random variables. The pdf can be calculated in closed form by calculating the characteristic function and then performing an inverse Fourier transform \citep{Imhof}, which generally produces unwieldy expressions, or through more recently discovered numerical methods \citep{Abhranil2021}. We present probability density functions for special cases relevant to 21-cm power spectrum estimation in this paper. A well-known result is that the modulus square of a circular Gaussian random variable is an exponential random variable i.e. one whose pdf is of exponential form:
\begin{equation}
    f_X(x) = \lambda e^{-\lambda x}.
\end{equation}
We refer to $\lambda$ as the ``scale parameter" of the exponential random variable. This relation can be gleaned by considering the complex Gaussian variable as a 2-d real Gaussian and inspecting the density in polar coordinates. Since it is instructive, and correlations are often short-ranging, we derive the sampling distribution for a sum of independent exponential random variables, and also show that in some instances, even sums of independent exponential random variables do not approach Gaussianity in the infinite limit. 

If $\bold{B}=\bold{I}$, and  each component of $\bold{X}$ is independent of every other component, but the variance of each component is not necessarily identical, then the variable Y,
\begin{equation}
    Y \equiv \bold{X}^\dag\bold{X}
\end{equation}
is called a ``hypoexponential" random variable. A special case of the hypoexponential random variable is the Erlang random variable, which occurs when all the variances of $\bold{X}$ are equal. The scale parameters for each of the terms in the sum, denoted $\lambda_j$, are equal to $1/\sigma_j^2$, where $\sigma_j^2$ is the total variance $j$th vector component of $\bold{X}$ (sum of variance of real and imaginary components). Due to the convolution theorem, the characteristic function of $Y$ is the product of the characteristic functions of each exponential term:
\begin{equation}
    E[e^{itY}] = \prod_{j=1}^N\frac{\lambda_j}{\lambda_j - it}
\end{equation}
where $N$ is the dimension of $\bold{X}$. Note that the variances do not all have to be distinct in this treatment. To compute its probability density function, we can perform an inverse Fourier transform:
\begin{equation}
    f_Y(y) = \frac{1}{2\pi}\int_{-\infty}^{+\infty} e^{-ity}\prod_{j=1}^N\frac{\lambda_j}{\lambda_j - it}dt.
\end{equation}
We can compute this integral using contour integration methods. The integrand is a meromorphic function i.e. it is complex differentiable everywhere except for a countable number of poles. It has one pole per distinct variance, and the order of each pole is given by the multiplicity of that variance. All of the poles are located in the lower half of the complex plane. When $y < 0$, we can use a semi-circular contour that encloses the upper-half plane and Jordan's lemma \citep{Byron1970}, showing that the result is 0. This is just an expression of positive-definiteness. When $y \geq 0$, we enclose the lower-half plane (and thus the residues) with a semi-circular contour. The residues are readily calculated using Cauchy's integral formula \citep{Byron1970}. The final result in the fully general case is
\begin{equation}
    f_Y(y) = \sum_{j=1}^n\sum_{m=0}^{m_j-1}\frac{\lambda_j^{m_j}}{(m_j-m-1)!m!}y^{m_j-m-1}e^{-\lambda_jy}\Psi_{mj}(-\lambda_j),
    \label{eq:fully_general}
\end{equation}
where $m_j$ is the multiplicity of the $j$th variance, and $\Psi_{mj}$ is the function
\begin{equation}
    \Psi_{mj}(t) = \frac{d^m}{dt^m}\bigg[\prod_{l\neq j}\bigg(\frac{\lambda_l}{\lambda_l+t}\bigg)^{m_l}\bigg].
\end{equation}
This is the probability density function for the positive-definite noise bias in a straight-square estimator when all the $\bold{k}$ modes are independent, up to scaling factors from the weights (and the fact that this represents a sum, not an average), which just rescale the $\lambda_j$'s.

The hypoexponential random variable is also a special case of random variable that arises in the study of finite-state continuous-time Markov chains, called a phase-type random variable \citep{Breuer-Baum}. A phase-type random variable describes the sampling distribution for a finite-state Markov process to reach an absorbing state. These processes are frequently studied in queuing theory, telecommunications theory, and biology \citep{Nelson1995, Breuer-Baum}. The hypoexponential distribution is the absorption time sampling distribution for a Markov process where all states are attainable and there are no cycles. Its distribution functions can be efficiently computed using the following theorem.
\begin{theorem}[\citet{Breuer-Baum} Theorem 9.2, Example 9.4]
The cumulative distribution function of the sum of independent, exponentially distributed random variables with scale parameters $\lambda_1,...,\lambda_n$ is equal to $1-\boldsymbol{\alpha}^Te^{\boldsymbol{\theta} x}\mathbf{1}$, and the pdf of the sum is $-\boldsymbol{\alpha}^Te^{\boldsymbol{\theta} x}\boldsymbol{\theta}\mathbf{1},$ where $\alpha\in \mathbf{R}^n$, $\boldsymbol{\alpha}=\begin{bmatrix} 1\\ 0 \\ 0 \\...\\0\end{bmatrix},$ $\mathbf{1}\in\mathbf{R}^n$ is $\mathbf{1}=\begin{bmatrix}1\\1\\...\\1\end{bmatrix},$ and $\boldsymbol{\theta} = \begin{bmatrix}-\lambda_1 & \lambda_1 &0&...&...&0\\
0 & -\lambda_2 & \lambda_2 & 0 &... & 0 \\
...& ... & ...& ...&...&...\\
0 & 0 &... & ...&-\lambda_{n-1} & \lambda_{n-1} \\
0 & 0 & ... & ...& 0 & -\lambda_n\end{bmatrix}$ is an $n\times n$ matrix. Here $e^\mathbf{A}=\sum_{i=0}^\infty \frac{\mathbf{A}^n}{n!}$ is the matrix exponential.\footnote{The quantity $\boldsymbol{\alpha}^Te^{\boldsymbol{\theta} x}\mathbf{1}$ is independent of the ordering of the scale parameters $\lambda_1,...,\lambda_n,$ though  $e^{\boldsymbol{\theta} x}$ is not. This is most clearly understood through the connection between hypoexponential distributions and absorption-time distributions for Markov chains explored in detail in \citet{Breuer-Baum}.}
\end{theorem}
In the next subsection, we use this form of the hypoexponential random variable to efficiently compute the distribution functions of some pedagogical hypoexponential examples with a high degree of numerical stability. In particular, we expose some distinct cases in which the hypoexponential random variable is and is not approximately Gaussian.

\subsection{Numerical Exploration}
\label{sec:straight-square-sim}

\begin{figure}
    \centering
    \includegraphics[scale=0.5]{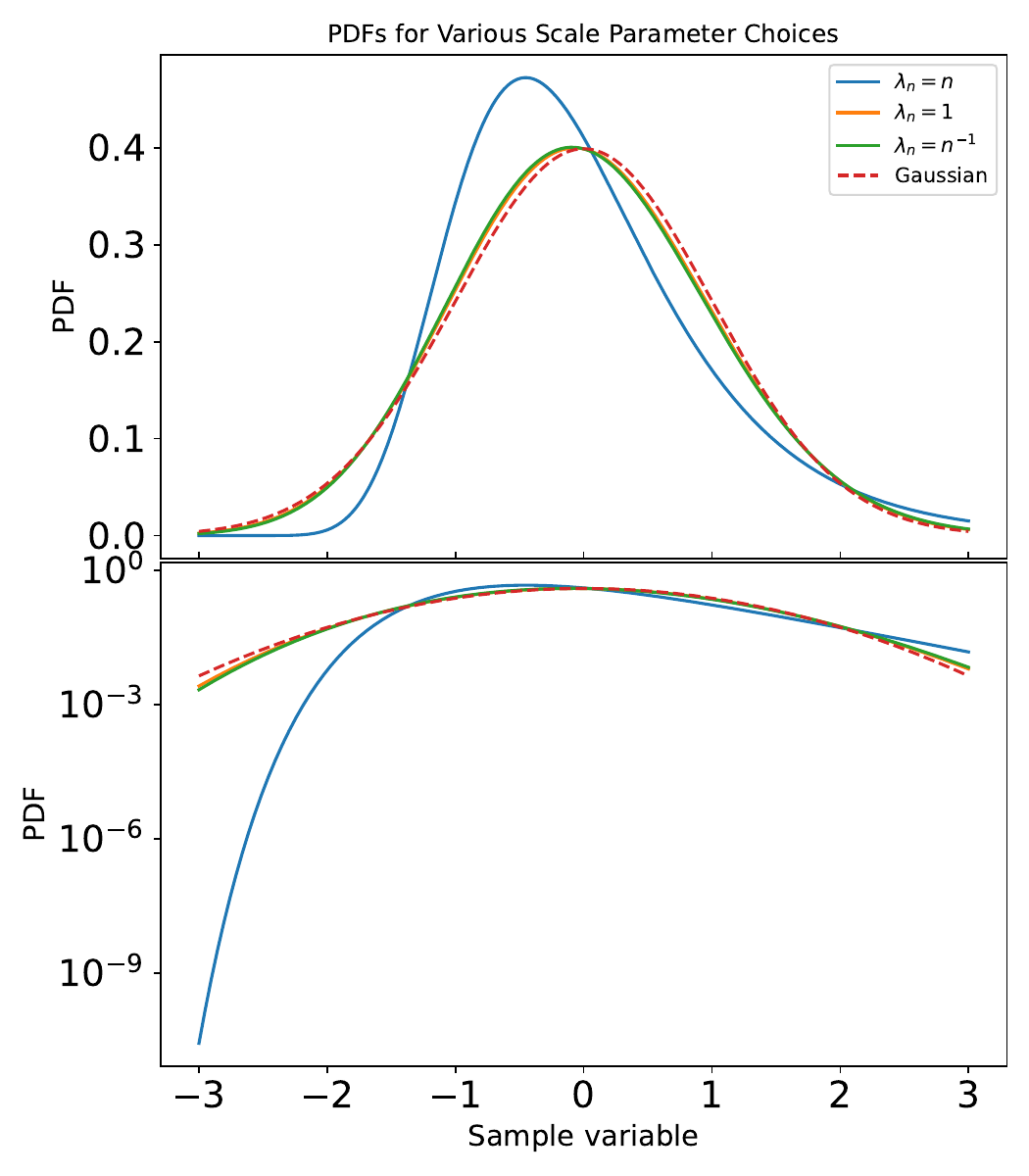}
    \caption{PDF plot for sums of 200 independent exponential random variables (one with linear and one with semilog axes), with scale parameters $\lambda_n$ for each sequence as indicated by the legend, compared to a Gaussian distribution. All variables are translated to have mean 0 and scaled to have variance 1. When $\lambda_n=n,$ the sum retains its skew as computed in Equation \ref{skew_integer_sum}.}
    \label{fig:norm_integer_reciprocal_fig_gauss_comp}
\end{figure}

Of particular note for hypoexponential random variables is that there are some sets of $\lambda_j$'s for which the variable is not even approximately Gaussian in the infinite limit. We can understand this by examining the skew of the hypoexponential random variable. Since the second and third central moments are cumulants (they add for independent random variables), we can formulate the skew of the hypoexponential as
\begin{equation}
E\bigg[\bigg(\frac{Y-\mu}{\sigma}\bigg)^3\bigg] = \frac{\sum_{j=1}^N\frac{2}{\lambda_j^3}}{\big(\sum_{j=1}^N\frac{1}{\lambda_j^2}\big)^{3/2}}.
\label{eq:skew}
\end{equation}
Since a Gaussian random variable has 0 skew, and we can construct sums in the form above that provide a nonzero skew in the infinite limit, we can say that some sums of exponential random variables do not converge to a Gaussian even in the infinite limit. To begin listing a few examples, consider when all the $\lambda_j$'s are identical. We then obtain a skew of $2/\sqrt{N}$, which vanishes in the infinite limit. If the $\lambda_j$'s are the reciprocals of the positive integers, then we also see the ratio of sums in Equation \ref{eq:skew} vanishes in the infinite limit. In contrast, if the $\lambda_j$'s are the positive integers, then the ratio of sums converges to a nonzero value. In particular it converges to the value
\begin{equation}\label{skew_integer_sum}
    \lim_{N \to \infty} \frac{\sum_{j=1}^N\frac{2}{\lambda_j^3}}{\big(\sum_{j=1}^N\frac{1}{\lambda_j^2}\big)^{3/2}} = \frac{2\zeta(3)}{\zeta(2)^{3/2}},
\end{equation}
where $\zeta(x)$ is the Riemann zeta function. We show a figure demonstrating these sums in Figure \ref{fig:norm_integer_reciprocal_fig_gauss_comp}, where we plot the probability density functions of these choices of hypoexponential random variables.
Note that this is not identical to the Lyapunov condition from \S\ref{sec:CLT}, so we cannot necessarily say that when the skew vanishes in the limit that the sum of exponentials converges to Gaussianity. However, it is clear from Figure \ref{fig:norm_integer_reciprocal_fig_gauss_comp} that some finite sums of exponential random variables are approximately Gaussian. 
%
%

Many error estimates fundamentally work with the estimated standard deviation of the power spectrum bin (typically reporting a $2\sigma$ upper limit based on the estimate of $\sigma$). For a Gaussian random variable, particular distances from the origin, measured in standard deviations, correspond uniquely to well-known values of the cdf of a Gaussian random variable. When the error distribution is non-Gaussian, reporting a significance in terms of a standard deviation no longer corresponds to the same values of the relevant cdf. For any hypoexponential random variable $X$, there is an exponential random variable $Y$ such that, for large $x$ $f_Y(x)<f_X(x)$, meaning that a given significance reported in standard deviations will tend to underestimate the size of the corresponding confidence interval if the significance is interpreted according to the correspondence between confidence intervals and standard deviations for a normal distribution. For nearly Gaussian hypoexponential random variables, the significance at which the difference in tails matters is high, i.e. one does not observe the non-Gaussianity except at distances far from the origin. Since many analyses report a $2\sigma$ upper limit, we characterize the degree to which this underestimates the actual inferred confidence interval as a figure of merit. If $\Phi(x)$ is the standard normal cumulative distribution function,
\begin{equation}
    \Phi(x) = \frac{1}{2}(1 + \text{erf}(x/\sqrt{2})),
\end{equation}
and $\text{erf}(x)$ is the error function,
\begin{equation}
    \text{erf}(x) = \frac{2}{\sqrt{\pi}}\int_0^{x}e^{-t^2}dt.
\end{equation}
then we calculate
\begin{equation}
    Q \equiv \frac{F^{-1}(\Phi(2)) - \mu}{2\sigma}
    \label{eq:Q_def}
\end{equation}
where $\mu$ is the mean of the variable in question. For a Gaussian random variable, $Q$ is exactly 1. For a single exponential random variable,
\begin{equation}
    Q_{\text{exp}} = \frac{-\ln (1 - \Phi(2)) - 1}{2} = 1.39159\text{...}
\end{equation}

This is to say that, for a \textit{single} exponential random variable, if one perfectly subtracts the noise bias and quotes a $2\sigma$ noise level based on an estimate of the standard deviation in order to represent a typical point at which a measurement might be designated as noise limited, then this will undershoot the actual 97.7\% noise level by approximately 40\%. Generally we observe that for any hypoexponential random variable, $Q$ is bounded between 1 and the value above. While this has the consequence that the power at which a measurement can be deemed noise limited is higher, it also means that the reported upper limits on the signal are inaccurate without reference to the true sampling distribution. Since any actual upper limit entails averaging multiple bins together, the amount of undershoot on the noise level is generally less than $Q_\text{exp}$ for a straight-square estimator, and decreases substantially with just a few averages ($N \sim 20$) assuming the noise in each sample is about the same. 

The exact inaccuracy of the upper limit calculation imposed by the non-Gaussianities will depend on the reported value as well as the particulars of the contributing voxels. Specifically, the upper limit is derived from the probability distribution of the signal, conditional on the power spectrum measurement. Neglecting signal-noise cross terms, Equation \ref{eq:straight_square} reads (omitting function arguments for brevity)
\begin{equation}
    P = |s|^2 + |n|^2.
\end{equation}
Since these two terms are positive definite, $|s|^2 \leq P$. This means that the measured value always acts as a 100\% upper limit on the signal. If the noise distribution is treated as Gaussian, then one may inadvertently report an erroneous ``$2\sigma$" upper limit greater than $P$ in some cases. This also suggests a potentially simpler method for placing upper limits when cross terms can be ignored, i.e. just quoting $P$ as the 100\% upper limit. However, this can make comparisons between analyses with very different sampling distributions difficult, such as the cross-multiply estimators in \S\ref{sec:cross-mult} which have an infinite 100\% upper limit.

We remark that $Q$ is only a proxy for the relative Gaussianity of the distribution. One could instead regard $Q$ as a function so that what we have defined as $Q$ is $Q(2)$. Then one may calculate $Q(x)$ for some $x$ greater than 2 if agreement with a Gaussian to some higher level of significance is desired. In other words, if the analyst only has reason to be concerned with a certain level of significance, $x$, then if $Q(x)$ is close to 1, a Gaussian approximation may suffice for that significance, but this depends on the context in which the error distribution is being used and so one should take care to understand the distribution to the extent that it affects their analysis, e.g. if one is performing inference based on the cosmological 21-cm signal.

\begin{figure}
    \centering
    \includegraphics[scale=0.45]{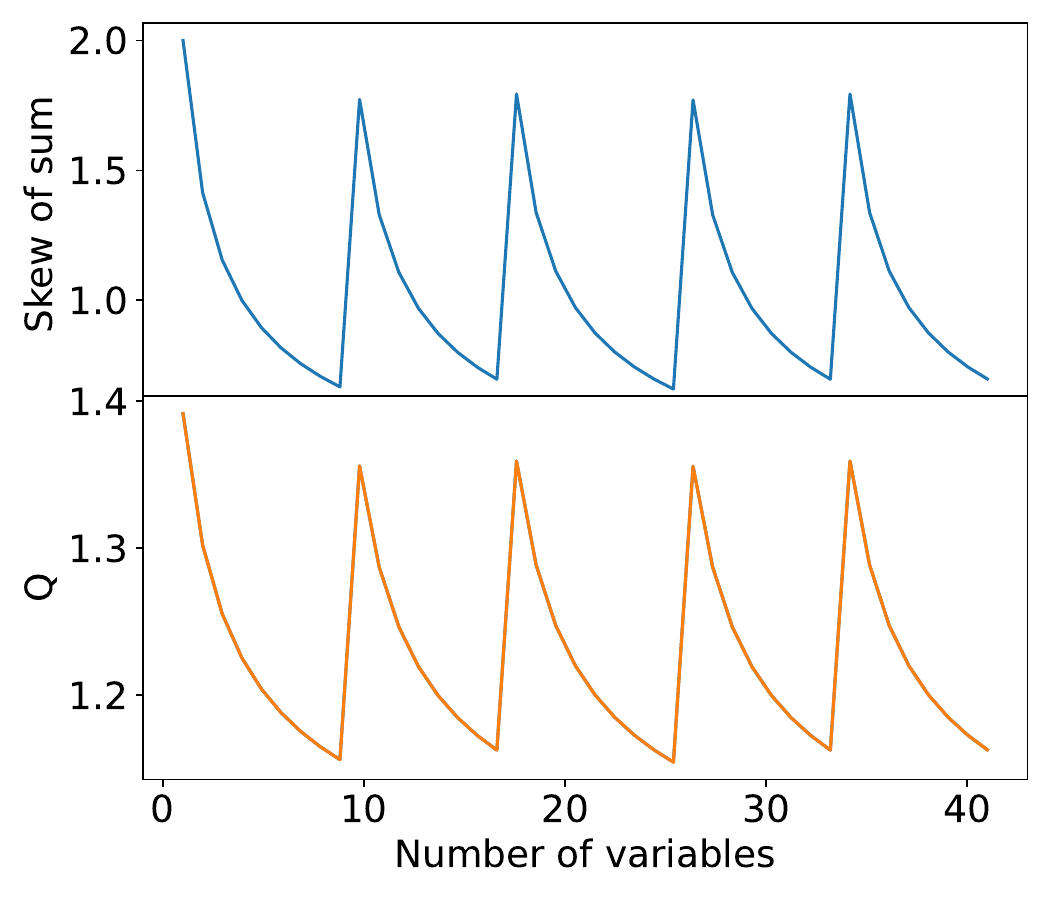}
    \caption{The skews and Q-values of sums of independent exponentials with scale parameters given by $10^{-\lfloor 5n/42\rfloor}$ for $n<42$. Each time an exponential is added with a scale parameter an order of magnitude smaller than the previous scale parameters, the tails revert to looking nearly exponential, as evidenced by the $Q-$value and skew spikes.}
    \label{fig:skews_and_qs}
\end{figure}
In order to intentionally create a pathological example for which the Lyapunov CLT does not apply, and to expose a type of behavior that can disrupt the Gaussianity of a random variable,  we add a sequence of independent, exponentially distributed random variables with scale parameters that decrease exponentially to $0$ in discrete steps. We add exponential random variables of a fixed scale parameter, but after some number of additions, we begin adding exponential random variables with a scale parameter a factor of 10 lower than the initial one. We repeat this with smaller and smaller scale parameters (i.e. larger and larger variances), decreasing by a factor of 10 each time. We see in Figure \ref{fig:skews_and_qs} that, in such cases, the distribution of the sum resembles the distribution of the exponential with the smallest scale parameter included in the sum. When several variables are added with a fixed scale parameter, we observe the sum beginning to converge to a Gaussian distribution, though in this ``comb'' example the dropping scale parameters prevent it from ever getting very close.

A unifying theme of the examples that do not converge to a Gaussian in the infinite limit is that the distribution of variances among the summed variables spans many scales (though this is not uniformly sufficient, e.g. the green curve in Figure \ref{fig:norm_integer_reciprocal_fig_gauss_comp}). Translating this to a practical application, analysts should be wary of combining measurements with substantially different noise levels. For example, Figure \ref{fig:skews_and_qs} demonstrates a scenario in which a sum that is tending towards Gaussianity can be shocked back towards exponentiality whenever a sample of standard deviation ten times the typical value is added.\footnote{To be clear, the final result is independent of the order of summation, but the orderliness of this example helps isolate the relevant phenomenon.} A more thorough examination of the error distribution is warranted when there is a high degree of variability in the uncertainty of co-added samples. We see this theme return through different examples in the next section, which discusses cross-multiply estimators. 

However, we also point out that most power spectrum estimators use ``inverse variance weighting," which in a straight-square estimator will weight each summed exponential variable by an estimate of its scale parameter (recall the scale parameter is the inverse variance of the sample to be squared). If this estimate is exact, then the noise-only term will be a sum of identically distributed exponential random variables, collapsing to the simplest statement of the central limit theorem if statistical dependence can be ignored. In general, the estimate will have errors due to imperfect knowledge of the system temperature or whatever other parameters guide the estimate. If the errors are small, then one could appeal to the Lyapunov CLT for intuition. We demonstrate this in the next section. If the errors are large, and in particular if they are much larger for some data than others, then a significantly non-Gaussian result may still be expected despite inverse variance weighting. 

\section{``Cross Multiply" Power Spectrum Estimators}
\label{sec:cross-mult}

Some power spectrum estimators, such as \eps, and that deployed in the recent HERA limits \citep{HERA2022, HERA2022C}, involve cross-multiplying data that share a signal component but have independent noise components. This results in a different error distribution, notably one that is not positive-definite but rather zero-mean. In the case of HERA, the estimator takes the form \citep{Tan2021},
\begin{multline}
    P(\bold{k}, t) = |s(\bold{k}, t)|^2 + s(\bold{k}, t)n_2^*(\bold{k}, t) + s^*(\bold{k}, t)n_1(\bold{k}, t) \\ + n_1(\bold{k}, t)n_2^*(\bold{k}, t)
\end{multline}
and similarly for the estimator in \eps$\text{ }$ \citep{Barry2019a}, except only involving the real component of this quantity. Here, $n_1(\bold{k}, t)$ and $n_2(\bold{k}, t)$ are independent circular Gaussian random vectors. As in \S\ref{sec:straight-square}, we will ignore the signal-noise cross terms for the reasons expressed there. The noise-only term can be expanded as (omitting the arguments for brevity),
\begin{multline}
     n_1n_2^* = \Re(n_1)\Re(n_2) + \Im(n_1)\Im(n_2) \\ + i(\Im(n_1)\Re(n_2) - \Im(n_2)\Re(n_1))
\end{multline}

Since by circularity the real and imaginary components of either one of the noise vectors are independent, identically distributed, and zero-mean, the components of the product are also independent, identically distributed, and zero-mean. Each component is a sum of two terms that are products of independent, 1-dimensional Gaussian random variables. Denoting any one of these terms as $W$, and writing the standard deviations of $n_1$ and $n_2$ as $\sigma_1$ and $\sigma_2$ respectively, we calculate its probability density function using the rule for products of independent random variables:
\begin{equation}
    f_{W}(w) = \frac{1}{2\pi\sigma_1\sigma_2}\int_{-\infty}^{\infty}\frac{\mathop{dx}}{|x|} e^{-x^2 /2\sigma_1^2}e^{-w^2 / 2x^2\sigma_2^2}
\end{equation}
We are free to rewrite this like so:
\begin{equation}
    f_{W}(w) = \frac{1}{\pi\sigma_1\sigma_2}\int_{0}^\infty\frac{\mathop{dx}}{x}e^{-\frac{|w|}{2\sigma_1\sigma_2}(e^{\ln(x^2\sigma_2/|w|\sigma_1)} + e^{\ln(|w|\sigma_1/x^2\sigma_2)})}
\end{equation}
Using the substitution $t = \ln(x^2\sigma_2 / |w|\sigma_1)$, we can write
\begin{equation}
    f_{W}(w) = \frac{1}{\pi\sigma_1\sigma_2}\int_0^\infty\mathop{dt}e^{-|w|\cosh t/\sigma_1\sigma_2} = \frac{K_0(|w|/ \sigma_1\sigma_2)}{\pi\sigma_1\sigma_2}
\end{equation}
where $K_0$ is a modified Bessel function of the second kind, and the final identity can be found using contour integration methods \citep{Watson1966}.

$W$ is a sum of two independent terms with exactly this distribution. The resulting sum is distributed according to the autoconvolution of the above distribution. Rather than compute this convolution directly, we instead evaluate the square of the characteristic function, which is the characteristic function of the autoconvolution by the convolution theorem. From \cite{abramowitz+stegun}, we have
\begin{equation}
    K_0(x) = \frac{1}{2}\int_{-\infty}^\infty\mathop{dk}\frac{\cos(kx)}{\sqrt{k^2 + 1}}
\end{equation}
Since $\sin(kx)$ is odd and $\sqrt{k^2 + 1}$ is even, we may write this as
\begin{equation}
    \frac{K_0(|x|)}{\pi} = \frac{1}{2\pi}\int_{-\infty}^\infty\mathop{dk}\frac{e^{-ikx}}{\sqrt{k^2 + 1}}
\end{equation}
where we have taken advantage of the fact that only the even part of $e^{-ikx}$ contributes to the integral in order to introduce the $|x|$ on the left-hand-side. Fourier inversion and the convolution theorem implies then that the characteristic function of $W$ is
\begin{equation}
    E[e^{ikw}] = \frac{1}{\sigma_1^2\sigma_2^2k^2 + 1}
    \label{eq:biexp_char}
\end{equation}
which is precisely the characteristic function of a Laplacian random variable centered on the origin with a scale parameter
\begin{equation}
    \lambda = \frac{1}{\sigma_1\sigma_2}
    \label{eq:bihyp_rate}
\end{equation} 
Invertibility of the Fourier transform finally implies that the noise-only terms in the aforementioned cross-multiply estimators have a Laplacian probability density function before any incoherent averaging. Incoherent averaging of independent power spectra will then have noise-only terms whose pdfs are that of sums of independent Laplacian random variables. 

\subsection{Numerical Exploration with Simulations}

In this section, we explore the properties of sums of Laplacian random variables using various simulations. We show examples for different mixtures of scale parameters. Some mixtures clearly tend towards Gaussianity as they are summed, while others do not. We find that mixtures whose scale parameters are concentrated around some value seem to tend towards Gaussianity.

Recall the figure of merit used in \S\ref{sec:straight-square-sim}, expressed by Equation \ref{eq:Q_def}. We again use this figure of merit in an exploration of the relative Gaussianity of a finite sum of Laplacian random variables. For a single Laplacian random variable, we have
\begin{equation}
    Q_\text{laplace} = -\frac{1}{2\sqrt{2}}\ln\bigg(2(1-\Phi(2))\bigg) = 1.09249...
\end{equation}
For a Normal random variable, this is equal to exactly 1. In general, a sum of Laplacian random variables will have a $Q$ between these two values. Note that $Q$ is only a proxy for the Gaussianity of the random variable unless one can prove that the convergence in distribution is uniform. It does, however, allow us to speak to the degree to which citing a variance underestimates thermal errors. The effect on the upper limit in this case is that it is generally underestimated when the Gaussian approximation is used. The degree of underestimation can vary depending on the value of the measurement. Since the signal-only term is clearly positive definite, the probability distribution of the signal conditional on the measurement must be truncated below 0 \citep[][Appendix A]{Li2019}. If the measured value is close to 0, then this can shift the mass of the signal distribution in an asymmetric way.    

An analytic formula of the pdf for the case where all the scale parameters are distinct is given in \citet{Tan2021}, and a more general one can be derived for when some of the scale parameters are degenerate. These formulas involve sums of products of many numbers (similar to Equation \ref{eq:fully_general}), and we found them to be prone to numerical instability when large numbers of variables are summed. We have therefore elected to do most of our exploration using semi-numerical techniques. Using the convolution theorem, we have (for a general continuous random variable)
\begin{equation}
    \int_{-x_0}^{x_0}f_X(x)dx = \frac{x_0}{\pi}\int_{-\infty}^{\infty}\text{sinc}(x_0t)\hat{f}_X(t)dt,
\end{equation}
where $\hat{f}_X$ is the characteristic function of the random variable in question and
\begin{equation}
    \text{sinc}(x) = \frac{\sin(x)}{x}
\end{equation}
Since our probability density function is always even, we may write\footnote{To see this, it may help to note that since $f$ is real, $\hat{f}$ is also even, as can be gleaned from Equation \ref{eq:biexp_char}.}
\begin{equation}
    \begin{aligned}
    F_Y(y) & = \frac{1}{2} + \int_{0}^{y}f_Y(y')dy' \\
    &=\frac{1}{2} +  \frac{x_0}{\pi}\int_{0}^{\infty}\text{sinc}(x_0t)\hat{f}_Y(t)dt,
    \label{eq:sinc_int}
\end{aligned}
\end{equation}
if $Y$ is a sum of Laplacian random variables and $y > 0$. To establish $Q$ for some $Y$ of interest, we evaluate Equation \ref{eq:sinc_int} numerically using Gaussian quadrature for a handful of $y$ values in the interval $[1, 1.1]$ (which contains $Q_\text{laplace}$), and then invert the CDF using a simple linear interpolation scheme. We find the type of interpolation does not change the result meaningfully. All of this is implemented in \textsc{Python} using the \textsc{numpy} and \textsc{scipy} libraries \citep{numpy, scipy}.

\begin{figure*}
    \centering
    \includegraphics{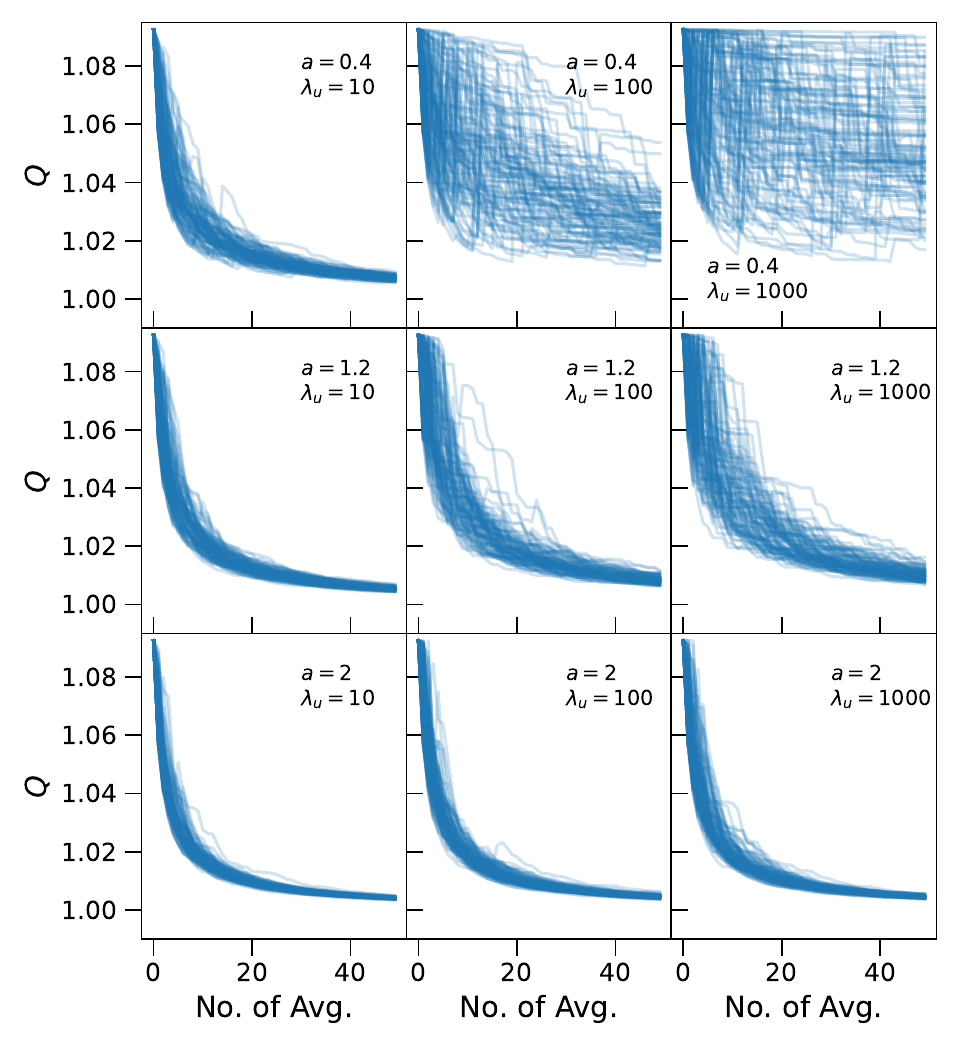}
    \caption{Ratio of the numerically calculated 97.7\% confidence interval divided by twice the standard deviation of sums of Laplacian random variables with different scale parameters (Equation \ref{eq:Q_def}). In each panel, the scale parameters for each line are drawn from a power law with index $a$, upper cutoff $\lambda_u$ (see annotations of each panel), and a lower cutoff of 1 (Equation \ref{eq:plaw_dens}). We simulate 100 independent groups of 50 scale parameters in each panel, and track the ratio of the confidence interval to twice the standard deviation as we cumulatively average the Laplacian random variables. This ratio is approximately 1.0925 for a Laplacian random variable. As the average proceeds towards Gaussianity, this ratio goes to 1. When scale parameters of very different size enter the average, the averaged variable returns closer to a Laplacian random variable than a Gaussian one. For power laws with large tails (upper panels), this effect is so pronounced that even after averaging 50 random variables, many of the distributions still appear Laplacian. On the other hand, steeper power laws are more concentrated towards the lower cutoff, and generally proceed towards Gaussianity in an orderly fashion (bottom panels). Note that we more often produce something approximately Gaussian than something significantly non-Gaussian.}
    \label{fig:plaw_fam}
\end{figure*}

In Figure \ref{fig:plaw_fam}, we calculate $Q$ for various mixtures of scale parameters. In each panel, we draw the scale parameters for a particular summed Laplacian from a power law with minimum cutoff at $\lambda=1$, and variable index and upper cutoff depending on the panel, i.e.
\begin{equation}
    f_\Lambda(\lambda) = \begin{cases}
    \frac{1-a}{\lambda_u^{1-a} - 1}\lambda^{-a}\text{, } 1 < \lambda < \lambda_u \\
    0\text{, otherwise}. 
    \end{cases}
    \label{eq:plaw_dens}
\end{equation}
We then calculate $Q$ for each partial sum over the mixture, making no special choice about the order of the scale parameters for each realization. For the case of inverse variance weighting, one should think of this as the distribution of the scale parameters after weighting the cross-multiplied samples, i.e. the distribution of scale parameters generated by errors in the estimates of the data variances. Each panel shows 100 different mixtures for the given power law parameters. We find that when the mixtures are less concentrated, which can be accomplished with shallower power law index and larger upper cutoff, the distribution of $Q$ at some given number of sums is also less concentrated. In the most diffuse case on display, even after 50 independent additions, $Q$ is more or less uniformly distributed over the possible range.

\begin{figure*}
    \centering
    \includegraphics{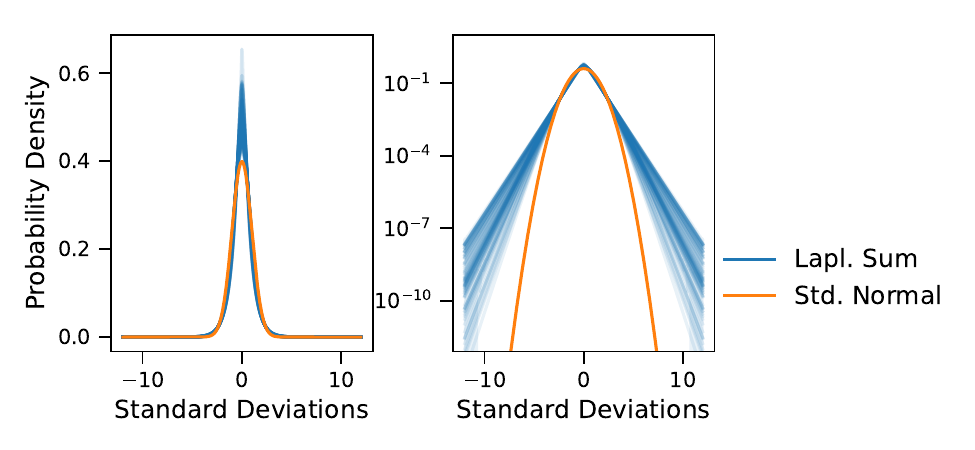}
    \caption{Collection of distributions of independent Laplacian sums after 50 additions, with scale parameters draw from a power law with index 0.4 and range 1 to 1000 (top right panel of Figure \ref{fig:plaw_fam}). Since the power law distribution of lambdas can result in wildly different final variances, each resulting variable has been standardized. We also plot a standard normal distribution for reference. The excess kurtosis is visible in realizations that resemble the original Laplacian distribution even after the summations. These are highly extreme cases that are unlikely to be encountered in practice. More realistic scale parameter distributions tend to produce more Gaussian results. }
    \label{fig:biexp_collect}
\end{figure*}

In Figure \ref{fig:biexp_collect}, we show the probability density functions of the summed Laplacians in the most extreme case of Figure \ref{fig:plaw_fam}, where the distribution of $Q$ after 50 additions is nearly uniform. We see that some of the PDFs demonstrate the kurtosis of the Laplacian, and some of them appear more Gaussian. We can also see there is significant variation in the rate of tail falloff on the semilog plot (right panel). In general, this is a rather extreme case of variance distribution for combined power spectrum measurements, and unlikely to be encountered in practice, since it essentially requires extremely uneven spatial mode or time sampling. 

\begin{figure*}
    \centering
    \includegraphics{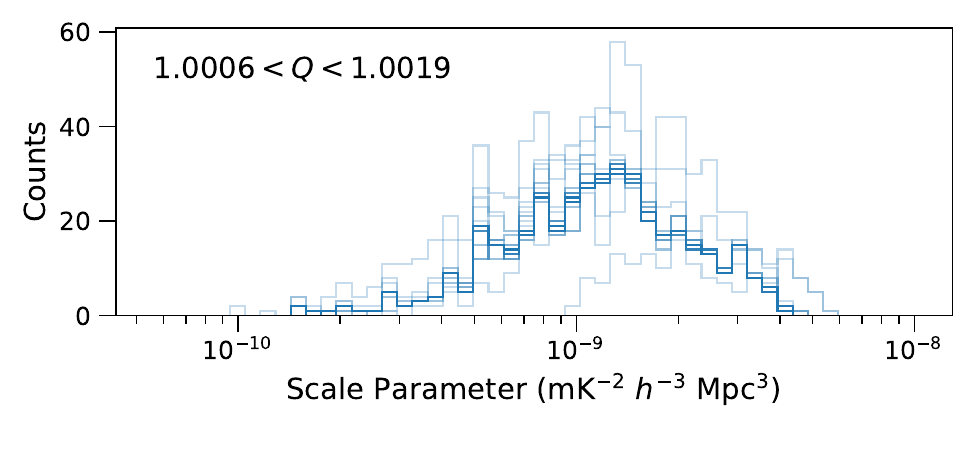}
    \caption{Histograms of scale parameters for 3d power spectrum bins estimated by $\varepsilon$\textsc{ppsilon} on one night of data, excluding data that contribute to the two lowest spherical $k$-modes. Each individual histogram (plotted with an opacity of 0.2 -- darker shades arise from overlap of multiple histograms) represents the scale parameters of the 3d bins that contribute to a given spherical wave number. The scale parameters are concentrated over approximately one order of magnitude. Calculating $Q$ from these scale parameters produces highly concentrated values close to 1, suggesting that each resulting error distribution would be highly Gaussian if statistical dependence between nearby voxels were negligible, and even if inverse variance weighting was not applied.}
    \label{fig:epps_hist}
\end{figure*}

In practice, error distributions are more likely to be Gaussian, e.g. \citet{Tan2021}. We can say from this analysis and the highly Gaussian error distributions in Figure 9 of that work that the power spectrum variance distributions for any spectra that were incoherently combined in \citet{HERA2022} were probably strongly concentrated. As another practical example, we show histograms of scale parameters that contribute to different spherical $k$-modes in the $\varepsilon$\textsc{ppsilon} estimator from one night of MWA data in Figure \ref{fig:epps_hist}. The scale parameters are concentrated over approximately one order of magnitude. When we calculate $Q$ using these scale parameters, we find that the values are extremely close to 1, indicating that there is negligible error in the uncertainty estimate by assuming a Gaussian distribution and quoting twice the standard deviation as the 97.7\% confidence interval. Here we have used all of the measured voxels that contribute to a spherical bin, including those that lie in the ``foreground wedge," a region of power spectrum space that is often excluded from spherical power spectrum averages due to the presence of immense foreground contamination \citep{Datta2010, Trott2012, Morales2012, Liu2014a}, which is unlike what is done in practice with MWA power spectra. Excluding modes with significant foreground contamination will use fewer bins, and therefore generally increase the $Q$-value in practice. However, this selection will not change the fact that the noise levels are generally homogeneous enough to produce approximately Gaussian sampling distributions, assuming enough independent samples are combined. Furthermore, $\varepsilon$\textsc{ppsilon} employs inverse variance weighting, which will homogenize the samples even more except in the unlikely event that the errors in the estimates are larger than the estimates themselves. One would need a model of the uncertainty in these particular estimates to investigate this fully.

Our analysis in this section as well as the previous one leads us to suggest that analysts should be cautious when incoherently combining power spectrum samples with vastly different statistics. We find that if variations of the scale parameters of the (potentially inverse variance weighted) power spectrum samples can be constrained to one order of magnitude, then approximate Gaussianity can be expected assuming more than a handful of samples are combined. About $N > 20$ appears sufficient in Figure \ref{fig:plaw_fam}, though the exact case should be evaluated by the analyst if there is any question. Including the occasional noisy sample does not totally prevent approximate Gaussianity. For example many of the $Q$-values are somewhat close to 1 in the bottom-right four panels of Figure \ref{fig:plaw_fam} after 50 additions, though shallower power laws are less reliable. We still urge caution for analysts using inverse variance weighting since estimates of the scale parameters for the noisiest samples may themselves be the estimates with the most error. While it is difficult to speak for every case, and we have yet to portray the effects of statistical dependence, we expect that practical circumstances will produce approximately Gaussian measurements of spherical power spectra more often than not, and so this assumption is usually justified.

\section{Dependent Random Variables}
\label{sec:dep_var}

Another central component to most basic statements of the central limit theorem are that the data be statistically independent. Complementary to the mathematical experiment depicted in Figure \ref{fig:norm_integer_reciprocal_fig_gauss_comp}, it is possible to combine statistically dependent but identically distributed data in such a way that Gaussianity is never achieved even in the infinite limit. An example of a potential occurrence of this in 21-cm power spectrum estimation is neatly described in Appendix A of \citet{Tan2021}, where it is shown that a persistent skew can be retained with a particular choice of delay spectrum estimator.  This is also an interesting example in the sense that the marginal distribution of each summand in the estimator is Laplacian, but due to the statistical dependency of the summands, a skew is produced.

Nevertheless, as discussed in \S\ref{sec:CLT}, there exist statements of the central limit theorem that relax the requirement that all the samples in question be independent. In particular, as long as the statistical dependency only has a finite range among the samples, then Gaussianity can be obtained in the infinite limit. In this section we explore how correlated thermal noise, which may exist at several stages of the pipeline in different forms depending on the particular estimator, behaves under averages. As an example of how correlations might arise in the thermal noise, consider that all 21-cm power spectrum estimation pipelines make use of a spectral tapering function in the frequency domain before Fourier transforming along this axis. This translates to a convolution in the dual space, meaning that samples which were originally uncorrelated in that space can become correlated. Dependency can also arise in gridded power spectrum estimators, since the information from a given visibility is usually assigned to several $uv$-pixels, meaning that there will be correlations in the perpendicular wave modes (which we ignored in the construction of Figure \ref{fig:epps_hist}). These correlations can produce nontrivial effects on the noise properties of spherically binned power spectra. The range of statistical dependency amongst the samples is highly idiosyncratic to the estimator in question, and a fully general treatment is beyond the scope of this paper. We therefore encourage analysts to understand the range of statistical dependencies in their data set and how they interplay to provide the ultimate sampling distribution of their estimate. We explore a particularly well-behaved and practical example next, in order to demonstrate the possibility of Gaussianity.

\subsection{Exploration Using a Fringe Rate Filter}

As a toy example, we explore the use of fringe rate filters \citep{Parsons2016}, which have been deployed to some extent in the HERA pipeline \citep{HERA2022, HERA2022C} and may be useful for further improvement of power spectrum upper limits. In short, a fringe rate filter can be implemented as a finite impulse response filter in time. One convolves a particular weighting function with the visibilities as a function of time, where the weighting function is chosen to enhance sensitivity to certain fringe rates (Fourier dual to the time axis of the data). Resulting samples will have some correlation length in time depending on the shape of the filter. Depending on how these correlated samples are used, these correlations can track all the way to the power spectrum estimate. If power spectra from correlated times are averaged incoherently, then depending on the strength of the correlation, the resulting error distribution may appear more or less Gaussian. For example, if one were to average perfectly correlated samples with one another, the sampling distribution would be unchanging at any length of averaging. This is an impractical case, since one would never intentionally generate a sequence of perfectly correlated data to average together, however it suggests that if one does produce strong correlations, then one may find that the error distribution only slowly migrates towards Gaussianity, if at all. 

We explore this possibility by implementing a fringe rate filter on a dynamic spectrum of Gaussian noise that is uncorrelated in both the frequency and time direction. To be clear, the following implementation of a fringe-rate filter and power spectrum estimation is not an exact replication of what is used by HERA or any other telescope currently. It is a stripped-down version that is meant to expose the effect of averaging correlated data on the 21-cm error distribution. We implement a naive bandpass filter in fringe-rate space that simply multiplies included modes by 1 and multiplies all other modes by 0. We express this in terms of linear algebra as follows. Let $\bold{n}_i$ be the noise samples for the $i$th baseline. Let $\bold{T}$ be the operator that performs the timelike Fourier transform. Then let $\bold{D}$ be a diagonal matrix that has a value of 1 for all included modes and 0 for all excluded modes. The fringe-rate filtration operator is then
\begin{equation}
    \bold{F} \equiv \bold{T}^{-1}\bold{D}\bold{T}.
\end{equation}
When the number of fringe rates modes is an odd number, this operator has a tidy closed form solution. The $jk$th component of $\bold{F}$ is given by
\begin{equation}
    F_{jk} = \frac{e^{2\pi i f_c (t_j - t_k)}}{N_t}\frac{\sin \big[(2N_f + 1)\pi\Delta f (t_j - t_k)\big]}{\sin \big[\pi\Delta f (t_j - t_k) \big]},
    \label{eq:D_N}
\end{equation}
where $N_t$ is the total number of times in the samples to be filtered, $f_c$ is the central fringe rate to be included, $t_j$ is the $j$th integration time in consideration, $2N_f+1$ is the total number of fringe rate modes being included (i.e. $N_f$ is the number of modes to either side of the central fringe rate being included), and $\Delta f$ is the fringe-rate spacing produced by the usual discrete Fourier transform conventions. This is sometimes referred to as the Dirichlet kernel, and looks like a periodic sinc function \citep{stein2003fourier}.

In any case, the fringe-rate filtered noise, $\tilde{\bold{n}}_i$, is given by
\begin{equation}
    \tilde{\bold{n}}_i = \bold{F}\bold{n}_i.
\end{equation}
Let $\bold{N}_i$ be the corresponding noise covariance matrix. Then the covariance matrix after fringe rate filtering, $\tilde{\bold{N}}_i$, is
\begin{equation}
    \tilde{\bold{N}}_i = E\big[\bold{F}\bold{n}_i\bold{n}_i^\dag\bold{F}^\dag \big] = \bold{F}\bold{N}_i\bold{F}^\dag.
\end{equation}

In the special case of uncorrelated Gaussian noise, where $\bold{N}_i$ is proportional to the identity (say with diagonal element, $\sigma_i^2$), we have
\begin{equation}
\begin{aligned}
    \tilde{\bold{N}}_i &= \sigma_i^2 \bold{F}\bold{F}^\dag \\ &= \sigma_i^2 \bold{T}^{-1}\bold{D}^2\bold{T} \\ &=
    \sigma_i^2\bold{T}^{-1}\bold{D}\bold{T} \\ 
    &= \sigma_i^2\bold{F}
\end{aligned}
\end{equation}
where we have canceled inverses where appropriate and made use of the fact that $\bold{D}$ is Hermitian and $\bold{D}^2=\bold{D}$. Given Equation \ref{eq:D_N}, we expect oscillating correlations and anticorrelations depending on the number of modes included, even when the original noise covariance is proportional to the identity.

If we want to include the effects of coherent averaging, we may make use of the fact that the covariance is a bilinear function of zero-mean random variables. Namely, if 
\begin{equation}
    Z = \frac{1}{N}\sum_{i=1}^Nz_i
\end{equation}
and
\begin{equation}
    W = \frac{1}{N}\sum_{i=1}^Nw_i
\end{equation}
and furthermore, the covariance of any $z_i$ with any other $z_j$ or $w_k$ is known, then (assuming all variables are zero-mean for notational simplicity)
\begin{equation}
    E\big[ ZW\big] = \frac{1}{N^2}\sum_{i=1}^N\sum_{j=1}^NE\big[ z_iw_j\big],
\end{equation}
and
\begin{equation}
    E\big[ Z^2\big] = \frac{1}{N^2}\sum_{i=1}^N\sum_{j=1}^NE\big[ z_iz_j\big],
\end{equation}
and similarly for $E\big[W^2\big]$. In other words, we may analytically produce the covariance matrix for coherent averaging of the fringe-rate filtered noise samples simply by summing the appropriate blocks of $\tilde{\bold{N}}_i$, and applying the $1/N^2$ factor. In the case of a weighted average, the weights must be tracked through the sums, and the $1/N^2$ factor changes to a quadratic function of the weights, but the overall process is the same. This allows us to short-circuit the simulation process. Rather than simulate independent samples and apply the fringe rate filter, we can instead generate jointly Gaussian samples that are consistent with the analytically propagated covariance matrix. 

From here, we can form a (noise) delay spectrum estimate, $\bold{p}_\tau$, by performing a delay transform (denoted by the operator $\bold{P}_\tau$) on two independently simulated fringe-rate filtered dynamic spectra and cross multiplying, in effect simulating the cross-multiplication of the delay spectra of two independent baselines at each (potentially coherently averaged) time bin:
\begin{equation}
   \bold{p}_\tau  \equiv  (\bold{P}_\tau \tilde{\bold{x}}_i)^*\circ  \bold{P}_\tau\tilde{\bold{x}}_j.
   \label{eq:ds_est}
\end{equation}
where $\circ$ is an elementwise-multiply. At this stage, the delay spectrum estimate (which we denote as vector-valued since we have computed one for each time at the specified delay mode) will be a Laplacian random variable at each time. The correlation between different times can be tracked using a relationship between the fourth and second moments of the $\tilde{\bold{x}}$. For real jointly Gaussian random vecrtors, this is known as Isserlis' theorem. It is generalized in \citet{Janssen1988} for complex jointly Gaussian random vectors (no circularity required). For white noise and two totally independent baselines, the answer is\footnote{It is common practice when delay transforming to apply a spectral tapering function, in which case $N_\nu$ gets replaced with the sum of the squared values of the tapering function at each frequency. There will also then be correlations between different delay modes, which should be accounted for if combining delay modes in a spherical average.}
\begin{equation}
    E\big[\bold{p}_\tau \bold{p}_\tau^\dag\big] = N_\nu^2 \tilde{\bold{N}}_i^*\circ\tilde{\bold{N}}_j
\end{equation}
where $N_\nu$ is the number of frequencies involved in the data. 

\begin{figure*}
    \centering
    \includegraphics{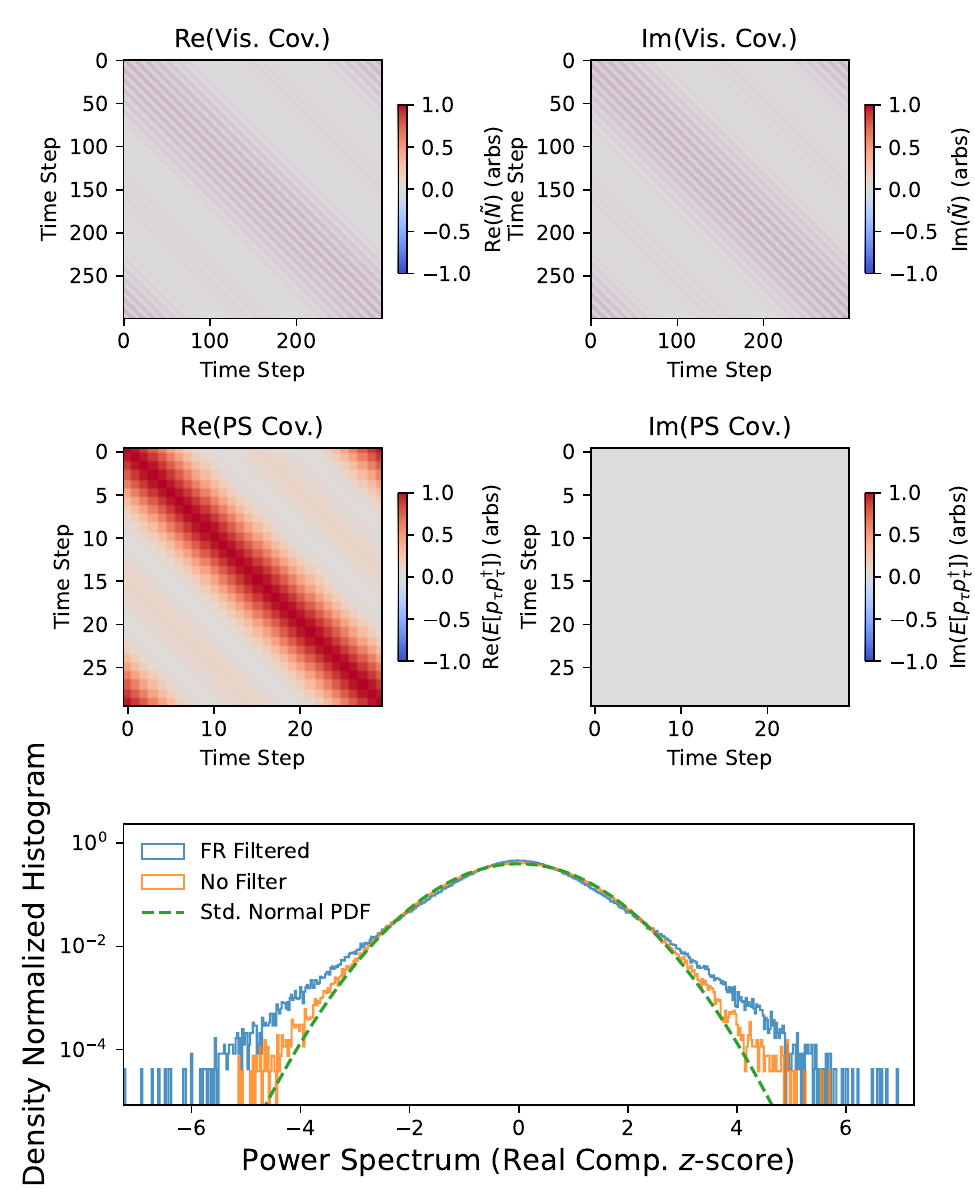}
    \caption{Top: Time-time covariance of independent and identically distributed noise after applying the DFT-based fringe rate filter described in the main text. In this case, only 3 fringe rates out of 150 are included (1\% of the available fringe modes), representing a worst case scenario in terms of induced correlation structure. Due to the DFT implementation of the filter, the correlation rises again at long lag, which could be avoided by e.g. convolving with discrete prolate spheroidal sequences \citep{Slepian1978, Ewall-Wice2021}. Middle: Time-time covariance of the delay spectra after fringe-rate filtering, coherently averaging over ten time steps, and cross multiplying totally independent baselines with identical noise distribution. Since the baselines are identical, the imaginary part of the covariance matrix is exactly 0. Bottom: Histogram of simulated delay spectra (real component) with fringe-rate filtering (blue and orange respectively). We normalize each datum according to its analytic standard deviation, and plot a Standard Normal pdf for reference (dashed green). In this case, where 30 time steps have been incoherently averaged, the distribution of the delay spectra is clearly non-Gaussian when the filter is applied. Conversely, the unfiltered samples agree reasonably well with the standard normal pdf until the more extreme depths of the tails.}
    \label{fig:delay_corr}
\end{figure*}

\begin{figure*}
    \centering
    \includegraphics{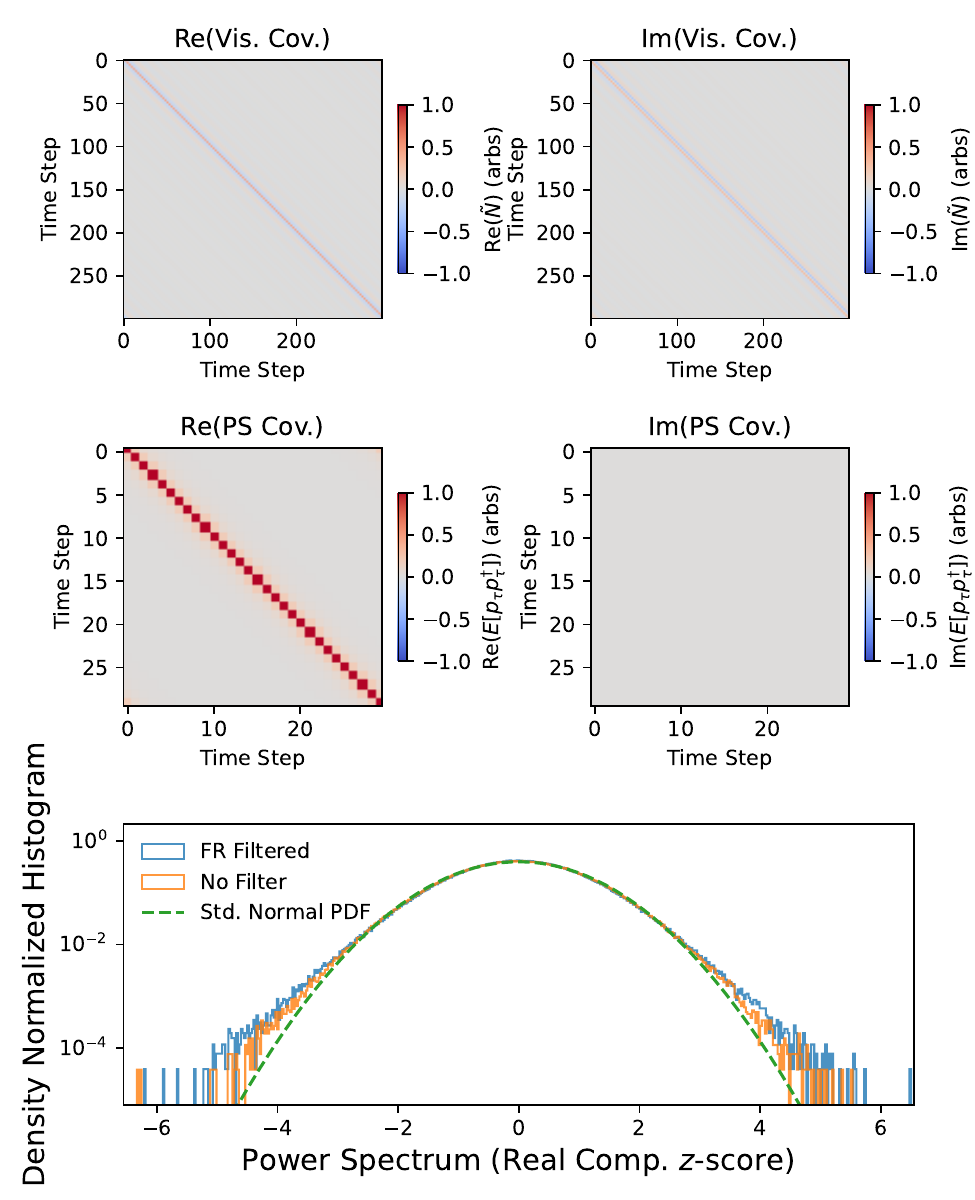}
    \caption{Same as Figure \ref{fig:delay_corr}, except including about 60 fringe modes out of 300 (20\% of the available modes). This noticeably shrinks the amount of correlation in the samples, and the resulting delay spectrum distribution appears nearly identical to the case in which no fringe-rate filtering was performed.}
    \label{fig:delay_corr2}
\end{figure*}

We are interested at this moment in investigating under what circumstances an incoherent average over the time axis of these estimates produces an approximately Gaussian distribution. To do this, we generate many samples of a circular Gaussian random vector according to the appropriate covariance matrix that represents the fringe-rate filtration and coherent averaging process in several different cases of each. We then form simulated power spectra from these samples according to Equation \ref{eq:ds_est}, incoherently average them, and histogram the results. We show this in Figures \ref{fig:delay_corr} and \ref{fig:delay_corr2} for different filter choices.

In each simulation, we start with 300 uncorrelated time samples. Since we can analytically propagate the covariance through the coherent averaging and Fourier transform steps, we directly sample from a circular multivariate Gaussian with consistent covariance as if we applied a fringe rate filter and coherently averaged 10 time steps (leaving 30 steps to be incoherently averaged later). In Figure \ref{fig:delay_corr}, we set $N_f=1$, which includes 3 modes out of the potential 150. This is an extremely narrow filter that is meant to represent the most extreme case of correlation structure that might be introduced by such a filter, and is not meant to represent a practical filter. In Figure \ref{fig:delay_corr2}, we use $N_f=30$ (about 20\% of the possible fringe modes), which is a potentially more realistic (though still quite narrow) setting depending on the goals of the analyst for the baselines in question. In each figure, we show the analytically calculated covariance matrix after fringe-rate filtering, as well as after coherent averaging and cross-multiplication with an independent baseline noise. We observe that for the extremely narrow filter, there is an induced correlation length in the delay spectra over approximately half of the samples in question. For the broader filter, the correlation length is about one sample to either side. The correlations extend around the time boundary due to the DFT implementation, which implicitly assumes a periodic input. These corner correlations could be avoided using an alternate implementation. The most similar alternative would be to apply a Toeplitz matrix made of discrete prolate spheroidal sequences \citep{Slepian1978, Ewall-Wice2021}, rather than the circulant matrix generated by our DFT implementation. The resulting delay spectra (from pure noise) are incoherently averaged, divided by their analytically calculated standard deviation, and histogrammed in the bottom panel. We also show a counterpart histogram with no fringe-rate filtration, and a standard normal pdf. 

For the extremely narrow filter (Figure \ref{fig:delay_corr}), there is a pronounced discrepancy between the tails of the filtered and unfiltered histograms. While the unfiltered samples are close to Gaussian throughout the range, the filtered samples significantly depart from the Gaussian pdf at about $3\sigma$. Both histograms appear to have exponential tails, meaning that the discrepancy worsens as we consider points further from the origin. While a comparison between the 97.7\% interval and $2\sigma$ may be relatively close, a figure of merit based further along the tail will inevitably produce strong discrepancies between significance in terms of the  actual CDF and the Gaussian equivalent. This may be particular important for inferences involving the cosmological 21-cm power spectrum signal, where the exact form of the likelihood of a model could drive the inference in a different direction. Fortunately for fringe rate filters, this effect appears to diminish rapidly as the size of the filter grows, as can be gleaned from Figure \ref{fig:delay_corr2}. However, due to the simplicity of this analysis (single baseline-pair, no spherical averaging, DFT implementation), and the potential ramifications for cosmological inference, we recommend that analysts assess the effects of potential correlations (or more general statistical dependence e.g. Appendix A of \citet{Tan2021}), in their own context.

\begin{figure*}
    \centering
    \includegraphics{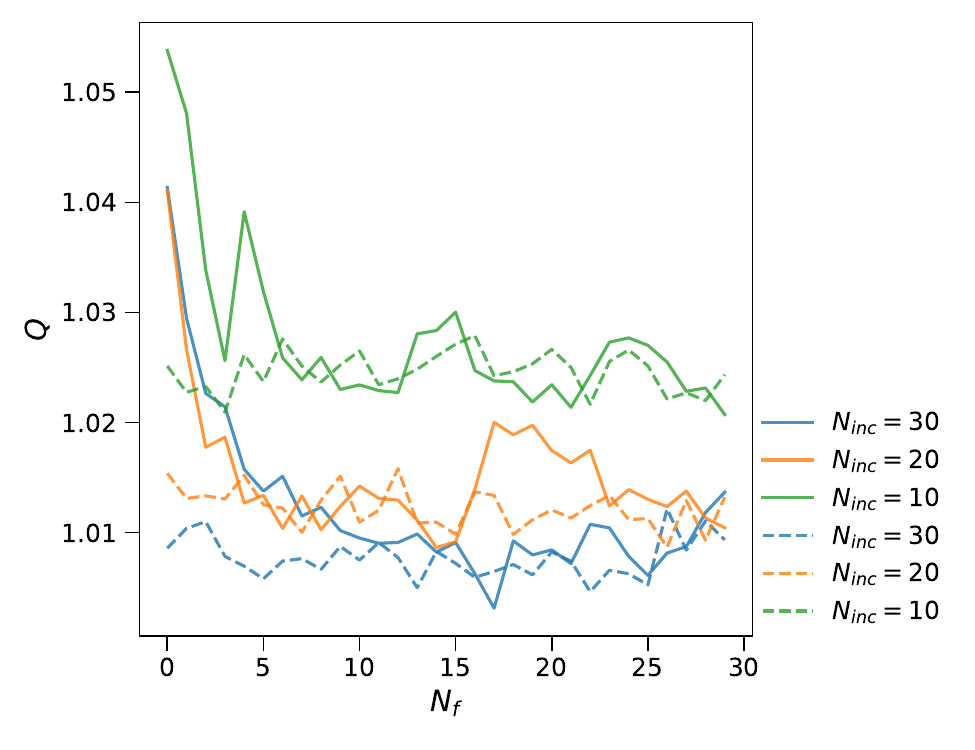}
    \caption{Ratio of 97.7\% confidence bound to twice the standard deviation for fringe rate filters of increasing size (solid), where the colors of the line indicate how many time steps were combined incoherently. We also plot non-filtered counterparts for each choice of incoherent averaging length (dashed). Results between the filtered and non-filtered counterparts are similar so long as the filter is sufficiently wide and enough spectra are combined incoherently. This plot suggests that so long as correlation lengths are relatively small in the data (a few samples at most), there is almost no difference between a treatment that takes these correlations into account and one that does not, though measures at further points on the tail may be more discrepant.}
    \label{fig:Q_corr}
\end{figure*}

Similar to analyses in previous sections, we find the ratio of the 97.7\% confidence interval to twice the standard deviation ($Q$-value; Equation \ref{eq:Q_def}) for the filter in different circumstances, shown in Figure \ref{fig:Q_corr}. In all cases, we consider 300 time steps, but within each realization we adjust the size of the filter or the coherent averaging length, thus adjusting the number of incoherent averages as well. We then construct delay spectrum samples, and statistically estimate $Q$ from them (solid lines). For each choice of incoherent averaging length, we also plot an estimate from unfiltered data but otherwise identical settings. We infer from the plot that this statistic quickly becomes limited by the incoherent averaging length rather than the correlation structure as we increase the breadth of the filter. It also suggests that all but the narrowest filters will produce noise properties that hardly alter the $2\sigma$ significance level. We stress again that this statistic does not speak for the entire distribution, and one should investigate as far along the tail as is relevant to one's analysis. However the bottom panel of Figure \ref{fig:delay_corr2}, which corresponds with the rightmost point of the blue line of Figure \ref{fig:Q_corr}, suggests that dramatic discrepancies in excess of what is observed in independent identically distributed noise may not appear until very high significance values when correlation lengths are less than a few samples.

\section{Summary and Conclusions}
\label{sec:conc}

In this work, we looked in detail at error distributions of common 21-cm power spectrum estimators. We focused on terms that only contained noise, ignoring cross terms with the cosmological and foreground signals. We proceeded by calculating the exact sampling distribution of the noise-only terms, and then investigating the effect of finite sums of random variables. The main guiding principle in this work was to understand when and why Gaussian approximations to the sampling distribution are appropriate. This is often attributed to the central limit theorem (CLT), despite the fact that the data in question do not obey the hypotheses of the most commonly known CLT, and despite the fact that all central limit theorems refer specifically to convergence in an infinite limit often with no reference to the effect of (finite) partial sums. Nevertheless, we found that in all but the most extreme violations of the typical hypotheses (namely independent and identically distributed samples), the finite sums indeed produced approximately Gaussian sampling distributions. This behavior is in congruence with a variety of CLTs that relax the classical hypotheses.

We broke power spectrum estimators into two groups: those involving direct squaring of the data in Fourier space, and those that cross-multiply independent samples. For a single sample, these have exponential and Laplacian sampling distributions, respectively. We then addressed the problem of understanding the summation of independent but non-identically distributed variables of the given type. We found that as long as the data are roughly homoschedastic (e.g. standard deviations spanning one order of magnitude), that one can typically expect an approximately Gaussian distribution by averaging a few dozen independent samples together. We also examined the violation of the independence assumption for the case of cross-multiplication estimators. We did this by implementing a fringe-rate filter on white noise and observing its effect on a single-baseline delay spectrum. While the treatment was highly simplistic, we expect that the lesson applies fairly generally, which is that approximate Gaussianity can still be expected as long as correlation lengths in the samples are less than a few samples and sufficiently many samples are averaged. However, long correlation lengths can produce distinctly non-Gaussian distributions. The detailed tracking of correlations and more general statistical dependencies among power spectrum samples is complicated and can appear in multiple stages of any given pipeline. We therefore recommend analysts investigate their own cases thoroughly.

Other than being an interesting mathematical exercise, this type of work is important for the scientific quality of upper limits on the cosmic reionization signal. Understanding the sampling distribution of one's estimator allows one to more accurately represent confidence intervals for the signal, and also allows one to write down more exact Bayesian likelihood functions. This ultimately impacts the quality of the statistical inference underlying our understanding of the astrophysical processes driving reionization.

\section{Ackowledgements}

This result is part of a project that has received funding from the European Research Council (ERC) under the European Union's Horizon 2020 research and innovation programme (Grant agreement No. 948764). This research was supported  NSF grant \#1613855. We gratefully acknowledge Phil Bull and Jianrong Tan for extremely helpful comments and discussions.

\section{Data Availability}

No new data were generated as a part of this work. The MWA data used is publicly available using the MWA All-Sky Virtual Observatory (ASVO).\footnote{\url{https://asvo.mwatelescope.org}}

\bibliographystyle{mnras}
\bibliography{main}

\begin{thebibliography}{}
\makeatletter
\relax
\def\mn@urlcharsother{\let\do\@makeother \do\$\do\&\do\#\do\^\do\_\do\%\do\~}
\def\mn@doi{\begingroup\mn@urlcharsother \@ifnextchar [ {\mn@doi@}
  {\mn@doi@[]}}
\def\mn@doi@[#1]#2{\def\@tempa{#1}\ifx\@tempa\@empty \href
  {http://dx.doi.org/#2} {doi:#2}\else \href {http://dx.doi.org/#2} {#1}\fi
  \endgroup}
\def\mn@eprint#1#2{\mn@eprint@#1:#2::\@nil}
\def\mn@eprint@arXiv#1{\href {http://arxiv.org/abs/#1} {{\tt arXiv:#1}}}
\def\mn@eprint@dblp#1{\href {http://dblp.uni-trier.de/rec/bibtex/#1.xml}
  {dblp:#1}}
\def\mn@eprint@#1:#2:#3:#4\@nil{\def\@tempa {#1}\def\@tempb {#2}\def\@tempc
  {#3}\ifx \@tempc \@empty \let \@tempc \@tempb \let \@tempb \@tempa \fi \ifx
  \@tempb \@empty \def\@tempb {arXiv}\fi \@ifundefined
  {mn@eprint@\@tempb}{\@tempb:\@tempc}{\expandafter \expandafter \csname
  mn@eprint@\@tempb\endcsname \expandafter{\@tempc}}}

\bibitem[\protect\citeauthoryear{Abramowitz \& Stegun}{Abramowitz \&
  Stegun}{1964}]{abramowitz+stegun}
Abramowitz M.,  Stegun I.~A.,  1964, Handbook of Mathematical Functions with
  Formulas, Graphs, and Mathematical Tables, ninth dover printing, tenth gpo
  printing edn.
Dover, New York

\bibitem[\protect\citeauthoryear{{Barry}, {Beardsley}, {Byrne}, {Hazelton},
  {Morales}, {Pober}  \& {Sullivan}}{{Barry} et~al.}{2019a}]{Barry2019a}
{Barry} N.,  {Beardsley} A.~P.,  {Byrne} R.,  {Hazelton} B.,  {Morales} M.~F.,
  {Pober} J.~C.,   {Sullivan} I.,  2019a, \mn@doi [Publications of the
  Astronomical Society of Australia] {10.1017/pasa.2019.21}, \href
  {https://ui.adsabs.harvard.edu/abs/2019PASA...36...26B} {36, e026}

\bibitem[\protect\citeauthoryear{{Barry} et~al.,}{{Barry}
  et~al.}{2019b}]{Barry2019b}
{Barry} N.,  et~al., 2019b, \mn@doi [The Astrophysical Journal]
  {10.3847/1538-4357/ab40a8}, \href
  {https://ui.adsabs.harvard.edu/abs/2019ApJ...884....1B} {884, 1}

\bibitem[\protect\citeauthoryear{{Bennett} et~al.,}{{Bennett}
  et~al.}{2003}]{bennett2003}
{Bennett} C.~L.,  et~al., 2003, \mn@doi [\apjs] {10.1086/377252}, \href
  {https://ui.adsabs.harvard.edu/abs/2003ApJS..148...97B} {148, 97}

\bibitem[\protect\citeauthoryear{Billingsley}{Billingsley}{2012}]{billingsley2012probability}
Billingsley P.,  2012, Probability and Measure.
Wiley Series in Probability and Statistics, Wiley

\bibitem[\protect\citeauthoryear{Breuer \& Baum}{Breuer \&
  Baum}{2005}]{Breuer-Baum}
Breuer L.,  Baum D.,  2005, An Introduction to Queueing Theory and
  Matrix-Analytic Methods.
Springer

\bibitem[\protect\citeauthoryear{{Byrne}, {Morales}, {Hazelton}  \&
  {Wilensky}}{{Byrne} et~al.}{2021}]{Byrne2021}
{Byrne} R.,  {Morales} M.~F.,  {Hazelton} B.~J.,   {Wilensky} M.,  2021,
  \mn@doi [\mnras] {10.1093/mnras/stab647}, \href
  {https://ui.adsabs.harvard.edu/abs/2021MNRAS.503.2457B} {503, 2457}

\bibitem[\protect\citeauthoryear{{Byron Jr.} \& {Fuller}}{{Byron Jr.} \&
  {Fuller}}{1970}]{Byron1970}
{Byron Jr.} F.~W.,  {Fuller} R.~W.,  1970, Mathematics of Classical and Quantum
  Physics.
Dover Publications, Inc.

\bibitem[\protect\citeauthoryear{Das \& Geisler}{Das \&
  Geisler}{2021}]{Abhranil2021}
Das A.,  Geisler W.~S.,  2021, \mn@doi [Journal of Vision]
  {10.1167/jov.21.10.1}, 21, 1

\bibitem[\protect\citeauthoryear{{Datta}, {Bowman}  \& {Carilli}}{{Datta}
  et~al.}{2010}]{Datta2010}
{Datta} A.,  {Bowman} J.~D.,   {Carilli} C.~L.,  2010, \mn@doi [The
  Astrophysical Journal] {10.1088/0004-637X/724/1/526}, \href
  {https://ui.adsabs.harvard.edu/abs/2010ApJ...724..526D} {724, 526}

\bibitem[\protect\citeauthoryear{{DeBoer} et~al.,}{{DeBoer}
  et~al.}{2017}]{DeBoer2017}
{DeBoer} D.~R.,  et~al., 2017, \mn@doi [Publications of the Astronomical
  Society of the Pacific] {10.1088/1538-3873/129/974/045001}, \href
  {https://ui.adsabs.harvard.edu/abs/2017PASP..129d5001D} {129, 045001}

\bibitem[\protect\citeauthoryear{{Ewall-Wice} et~al.,}{{Ewall-Wice}
  et~al.}{2021}]{Ewall-Wice2021}
{Ewall-Wice} A.,  et~al., 2021, \mn@doi [\mnras] {10.1093/mnras/staa3293},
  \href {https://ui.adsabs.harvard.edu/abs/2021MNRAS.500.5195E} {500, 5195}

\bibitem[\protect\citeauthoryear{{Furlanetto}, {Oh}  \& {Briggs}}{{Furlanetto}
  et~al.}{2006}]{Furl2006}
{Furlanetto} S.~R.,  {Oh} S.~P.,   {Briggs} F.~H.,  2006, \mn@doi [Physics
  Reports] {10.1016/j.physrep.2006.08.002}, \href
  {https://ui.adsabs.harvard.edu/abs/2006PhR...433..181F} {433, 181}

\bibitem[\protect\citeauthoryear{Gallager}{Gallager}{2013}]{Gallager2013}
Gallager R.,  2013, Stochastic Processes: Theory for Applications,
  \mn@doi{10.1017/CBO9781139626514.
}

\bibitem[\protect\citeauthoryear{{HERA Collaboration} et~al.,}{{HERA
  Collaboration} et~al.}{2022}]{HERA2022}
{HERA Collaboration} et~al., 2022, \mn@doi [\apj] {10.3847/1538-4357/ac1c78},
  \href {https://ui.adsabs.harvard.edu/abs/2022ApJ...925..221A} {925, 221}

\bibitem[\protect\citeauthoryear{{HERA Collaboration} et~al.,}{{HERA
  Collaboration} et~al.}{2023}]{HERA2023}
{HERA Collaboration} et~al., 2023, \mn@doi [\apj] {10.3847/1538-4357/acaf50},
  \href {https://ui.adsabs.harvard.edu/abs/2023ApJ...945..124H} {945, 124}

\bibitem[\protect\citeauthoryear{{Harris} et~al.,}{{Harris}
  et~al.}{2020}]{numpy}
{Harris} C.~R.,  et~al., 2020, \mn@doi [\nat] {10.1038/s41586-020-2649-2},
  \href {https://ui.adsabs.harvard.edu/abs/2020Natur.585..357H} {585, 357}

\bibitem[\protect\citeauthoryear{Hoeffding \& Robbins}{Hoeffding \&
  Robbins}{1948}]{HoeffRobbins}
Hoeffding W.,  Robbins H.,  1948, Duke Mathematical Journal, 15, 773

\bibitem[\protect\citeauthoryear{Imhof}{Imhof}{1961}]{Imhof}
Imhof J.~P.,  1961, Biometrika, 48, 419

\bibitem[\protect\citeauthoryear{Janssen \& Stoica}{Janssen \&
  Stoica}{1988}]{Janssen1988}
Janssen P.,  Stoica P.,  1988, \mn@doi [IEEE Transactions on Automatic Control]
  {10.1109/9.1319}, 33, 867

\bibitem[\protect\citeauthoryear{{Kern} \& {Liu}}{{Kern} \&
  {Liu}}{2021}]{Kern2021}
{Kern} N.~S.,  {Liu} A.,  2021, \mn@doi [\mnras] {10.1093/mnras/staa3736},
  \href {https://ui.adsabs.harvard.edu/abs/2021MNRAS.501.1463K} {501, 1463}

\bibitem[\protect\citeauthoryear{{Li} et~al.,}{{Li} et~al.}{2019}]{Li2019}
{Li} W.,  et~al., 2019, \mn@doi [The Astrophysical Journal]
  {10.3847/1538-4357/ab55e4}, \href
  {https://ui.adsabs.harvard.edu/abs/2019ApJ...887..141L} {887, 141}

\bibitem[\protect\citeauthoryear{{Liu} \& {Shaw}}{{Liu} \&
  {Shaw}}{2020}]{Liu2020}
{Liu} A.,  {Shaw} J.~R.,  2020, \mn@doi [Publications of the Astronomical
  Society of the Pacific] {10.1088/1538-3873/ab5bfd}, \href
  {https://ui.adsabs.harvard.edu/abs/2020PASP..132f2001L} {132, 062001}

\bibitem[\protect\citeauthoryear{{Liu}, {Parsons}  \& {Trott}}{{Liu}
  et~al.}{2014}]{Liu2014a}
{Liu} A.,  {Parsons} A.~R.,   {Trott} C.~M.,  2014, \mn@doi [Physical Review D]
  {10.1103/PhysRevD.90.023018}, \href
  {https://ui.adsabs.harvard.edu/abs/2014PhRvD..90b3018L} {90, 023018}

\bibitem[\protect\citeauthoryear{{Mertens}, {Ghosh}  \& {Koopmans}}{{Mertens}
  et~al.}{2018}]{Mertens2018}
{Mertens} F.~G.,  {Ghosh} A.,   {Koopmans} L.~V.~E.,  2018, \mn@doi [\mnras]
  {10.1093/mnras/sty1207}, \href
  {https://ui.adsabs.harvard.edu/abs/2018MNRAS.478.3640M} {478, 3640}

\bibitem[\protect\citeauthoryear{{Mertens} et~al.,}{{Mertens}
  et~al.}{2020}]{Mertens2020}
{Mertens} F.~G.,  et~al., 2020, \mn@doi [Monthly Notices of the Royal
  Astronomical Society] {10.1093/mnras/staa327}, \href
  {https://ui.adsabs.harvard.edu/abs/2020MNRAS.493.1662M} {493, 1662}

\bibitem[\protect\citeauthoryear{{Morales} \& {Wyithe}}{{Morales} \&
  {Wyithe}}{2010}]{Morales2010}
{Morales} M.~F.,  {Wyithe} J. S.~B.,  2010, \mn@doi [Annual Review of Astronomy
  and Astrophysics] {10.1146/annurev-astro-081309-130936}, \href
  {https://ui.adsabs.harvard.edu/abs/2010ARA&A..48..127M} {48, 127}

\bibitem[\protect\citeauthoryear{{Morales}, {Hazelton}, {Sullivan}  \&
  {Beardsley}}{{Morales} et~al.}{2012}]{Morales2012}
{Morales} M.~F.,  {Hazelton} B.,  {Sullivan} I.,   {Beardsley} A.,  2012,
  \mn@doi [The Astrophysical Journal] {10.1088/0004-637X/752/2/137}, \href
  {https://ui.adsabs.harvard.edu/abs/2012ApJ...752..137M} {752, 137}

\bibitem[\protect\citeauthoryear{{Morales}, {Beardsley}, {Pober}, {Barry},
  {Hazelton}, {Jacobs}  \& {Sullivan}}{{Morales} et~al.}{2019}]{Morales2019}
{Morales} M.~F.,  {Beardsley} A.,  {Pober} J.,  {Barry} N.,  {Hazelton} B.,
  {Jacobs} D.,   {Sullivan} I.,  2019, \mn@doi [Monthly Notices of the Royal
  Astronomical Society] {10.1093/mnras/sty2844}, \href
  {https://ui.adsabs.harvard.edu/abs/2019MNRAS.483.2207M} {483, 2207}

\bibitem[\protect\citeauthoryear{{Nelson, Randolph}}{{Nelson,
  Randolph}}{1995}]{Nelson1995}
{Nelson, Randolph} 1995, {Probability, stochastic processes, and queueing
  theory: the mathematics of computer performance modeling}.
Springer-Verlag

\bibitem[\protect\citeauthoryear{{Offringa}, {Mertens}  \&
  {Koopmans}}{{Offringa} et~al.}{2019}]{Offringa2019a}
{Offringa} A.~R.,  {Mertens} F.,   {Koopmans} L.~V.~E.,  2019, \mn@doi [Monthly
  Notices of the Royal Astronomical Society] {10.1093/mnras/stz175}, \href
  {https://ui.adsabs.harvard.edu/abs/2019MNRAS.484.2866O} {484, 2866}

\bibitem[\protect\citeauthoryear{{Paciga} et~al.,}{{Paciga}
  et~al.}{2013}]{Paciga2013}
{Paciga} G.,  et~al., 2013, \mn@doi [\mnras] {10.1093/mnras/stt753}, \href
  {https://ui.adsabs.harvard.edu/abs/2013MNRAS.433..639P} {433, 639}

\bibitem[\protect\citeauthoryear{{Parsons} et~al.,}{{Parsons}
  et~al.}{2010}]{Parsons2010}
{Parsons} A.~R.,  et~al., 2010, \mn@doi [\aj] {10.1088/0004-6256/139/4/1468},
  \href {https://ui.adsabs.harvard.edu/abs/2010AJ....139.1468P} {139, 1468}

\bibitem[\protect\citeauthoryear{{Parsons}, {Pober}, {Aguirre}, {Carilli},
  {Jacobs}  \& {Moore}}{{Parsons} et~al.}{2012}]{Parsons2012}
{Parsons} A.~R.,  {Pober} J.~C.,  {Aguirre} J.~E.,  {Carilli} C.~L.,  {Jacobs}
  D.~C.,   {Moore} D.~F.,  2012, \mn@doi [The Astrophysical Journal]
  {10.1088/0004-637X/756/2/165}, \href
  {https://ui.adsabs.harvard.edu/abs/2012ApJ...756..165P} {756, 165}

\bibitem[\protect\citeauthoryear{{Parsons}, {Liu}, {Ali}  \& {Cheng}}{{Parsons}
  et~al.}{2016}]{Parsons2016}
{Parsons} A.~R.,  {Liu} A.,  {Ali} Z.~S.,   {Cheng} C.,  2016, \mn@doi [\apj]
  {10.3847/0004-637X/820/1/51}, \href
  {https://ui.adsabs.harvard.edu/abs/2016ApJ...820...51P} {820, 51}

\bibitem[\protect\citeauthoryear{{Patil} et~al.,}{{Patil}
  et~al.}{2017}]{Patil2017}
{Patil} A.~H.,  et~al., 2017, \mn@doi [\apj] {10.3847/1538-4357/aa63e7}, \href
  {https://ui.adsabs.harvard.edu/abs/2017ApJ...838...65P} {838, 65}

\bibitem[\protect\citeauthoryear{Shevtsova}{Shevtsova}{2010}]{Shevtsova2021}
Shevtsova I.,  2010, Doklady Mathematics, 435, 862

\bibitem[\protect\citeauthoryear{{Slepian}}{{Slepian}}{1978}]{Slepian1978}
{Slepian} D.,  1978, AT T Technical Journal, \href
  {https://ui.adsabs.harvard.edu/abs/1978ATTTJ..57.1371S} {57, 1371}

\bibitem[\protect\citeauthoryear{Stein \& Shakarchi}{Stein \&
  Shakarchi}{2003}]{stein2003fourier}
Stein E.,  Shakarchi R.,  2003, Fourier Analysis: An Introduction.
Princeton University Press, \url
  {https://books.google.com/books?id=I6CJngEACAAJ}

\bibitem[\protect\citeauthoryear{{Tan} et~al.,}{{Tan} et~al.}{2021}]{Tan2021}
{Tan} J.,  et~al., 2021, \mn@doi [\apjs] {10.3847/1538-4365/ac0533}, \href
  {https://ui.adsabs.harvard.edu/abs/2021ApJS..255...26T} {255, 26}

\bibitem[\protect\citeauthoryear{{Thompson}, {Moran}  \& {Swenson}}{{Thompson}
  et~al.}{2017}]{Thompson2017}
{Thompson} A.~R.,  {Moran} J.~M.,   {Swenson} George~W. J.,  2017,
  {Interferometry and Synthesis in Radio Astronomy, 3rd Edition},
  \mn@doi{10.1007/978-3-319-44431-4.
}

\bibitem[\protect\citeauthoryear{{Tingay} et~al.,}{{Tingay}
  et~al.}{2013}]{Tingay2013}
{Tingay} S.~J.,  et~al., 2013, \mn@doi [\pasa] {10.1017/pasa.2012.007}, \href
  {Publications of the Astronomical Society of Australia} {30, e007}

\bibitem[\protect\citeauthoryear{{Trott}, {Wayth}  \& {Tingay}}{{Trott}
  et~al.}{2012}]{Trott2012}
{Trott} C.~M.,  {Wayth} R.~B.,   {Tingay} S.~J.,  2012, \mn@doi [The
  Astrophysical Journal] {10.1088/0004-637X/757/1/101}, \href
  {https://ui.adsabs.harvard.edu/abs/2012ApJ...757..101T} {757, 101}

\bibitem[\protect\citeauthoryear{{Trott} et~al.,}{{Trott}
  et~al.}{2016}]{Trott2016b}
{Trott} C.~M.,  et~al., 2016, \mn@doi [\apj] {10.3847/0004-637X/818/2/139},
  \href {https://ui.adsabs.harvard.edu/abs/2016ApJ...818..139T} {818, 139}

\bibitem[\protect\citeauthoryear{{VanderPlas}}{{VanderPlas}}{2018}]{VanderPlas2018}
{VanderPlas} J.~T.,  2018, \mn@doi [The Astrophysical Journal Supplement
  Series] {10.3847/1538-4365/aab766}, \href
  {https://ui.adsabs.harvard.edu/abs/2018ApJS..236...16V} {236, 16}

\bibitem[\protect\citeauthoryear{Virtanen et~al.,}{Virtanen
  et~al.}{2020}]{scipy}
Virtanen P.,  et~al., 2020, \mn@doi [Nature Methods]
  {10.1038/s41592-019-0686-2}, \href {https://rdcu.be/b08Wh} {17, 261}

\bibitem[\protect\citeauthoryear{{Watson, G.N.}}{{Watson,
  G.N.}}{1966}]{Watson1966}
{Watson, G.N.} 1966, {A Treatise on the Theory of Bessel Functions}

\bibitem[\protect\citeauthoryear{{Wayth} et~al.,}{{Wayth}
  et~al.}{2018}]{Wayth2018}
{Wayth} R.~B.,  et~al., 2018, \mn@doi [Publications of the Astronomical Society
  of Australia] {10.1017/pasa.2018.37}, \href
  {https://ui.adsabs.harvard.edu/abs/2018PASA...35...33W} {35, 33}

\bibitem[\protect\citeauthoryear{{Wilensky}, {Morales}, {Hazelton}, {Barry},
  {Byrne}  \& {Roy}}{{Wilensky} et~al.}{2019}]{Wilensky2019}
{Wilensky} M.~J.,  {Morales} M.~F.,  {Hazelton} B.~J.,  {Barry} N.,  {Byrne}
  R.,   {Roy} S.,  2019, \mn@doi [Publications of the Astronomical Society of
  the Pacific] {10.1088/1538-3873/ab3cad}, \href
  {https://ui.adsabs.harvard.edu/abs/2019PASP..131k4507W} {131, 114507}

\bibitem[\protect\citeauthoryear{{Wilensky}, {Hazelton}  \&
  {Morales}}{{Wilensky} et~al.}{2022}]{Wilensky2022}
{Wilensky} M.~J.,  {Hazelton} B.~J.,   {Morales} M.~F.,  2022, \mn@doi [\mnras]
  {10.1093/mnras/stab3456}, \href
  {https://ui.adsabs.harvard.edu/abs/2022MNRAS.510.5023W} {510, 5023}

\bibitem[\protect\citeauthoryear{{de Oliveira-Costa}, {Tegmark}, {Gaensler},
  {Jonas}, {Landecker}  \& {Reich}}{{de Oliveira-Costa}
  et~al.}{2008}]{deOliveira-Costa2008}
{de Oliveira-Costa} A.,  {Tegmark} M.,  {Gaensler} B.~M.,  {Jonas} J.,
  {Landecker} T.~L.,   {Reich} P.,  2008, \mn@doi [\mnras]
  {10.1111/j.1365-2966.2008.13376.x}, \href
  {https://ui.adsabs.harvard.edu/abs/2008MNRAS.388..247D} {388, 247}

\bibitem[\protect\citeauthoryear{{van Haarlem} et~al.,}{{van Haarlem}
  et~al.}{2013}]{vanHaarlem2013}
{van Haarlem} M.~P.,  et~al., 2013, \mn@doi [Astronomy and Astrophysics]
  {10.1051/0004-6361/201220873}, \href
  {https://ui.adsabs.harvard.edu/abs/2013A&A...556A...2V} {556, A2}

\makeatother
\end{thebibliography}

\end{document}